\documentclass[12pt]{iopart}
\usepackage{graphics}
\usepackage{url}
\usepackage[numbers,square,sort&compress]{natbib}

\begin{document}
  
\title[Spectral classification and 
EoS constraints in binary neutron star mergers]{Spectral classification of gravitational-wave emission and equation of state constraints in binary neutron star mergers}

\author{A. Bauswein$^{1,2}$ \& N. Stergioulas$^{3}$}

\address{$^{1}$GSI Helmholtzzentrum f\"ur Schwerionenforschung, 
Planckstra{\ss}e 1, 64291 Darmstadt, Germany}
\address{$^{2}$Heidelberg Institute for Theoretical Studies, 
Schloss-Wolfsbrunnenweg 35, 69118 Heidelberg, Germany}
\address{$^{3}$Department of Physics, Aristotle University of Thessaloniki, 54124 Thessaloniki, Greece}

\ead{a.bauswein@gsi.de, niksterg@auth.gr}
\vspace{10pt}
\begin{indented}
\item[]December 2018
\end{indented}

\begin{abstract}
In summer 2017 gravitational waves from a binary neutron star merger were detected for the first time. Moreover, electromagnetic emission was observed and associated with the merger. This very first unambiguous observation of a neutron star coalescence has impressively advanced our understanding of the merger process and has set some first constraints on the macroscopic properties of neutron stars, with direct implications for the high-density equation of state.
We discuss work on neutron star mergers focusing on the postmerger gravitational-wave emission. These studies are based on numerical simulations of the merger and survey a large sample of candidate equations of state for neutron star matter. The goal is to connect observables with the underlying physics questions. This offers a way to constrain the properties of high-density matter through the determination of neutron star radii, as inferred by an empirical relation connecting the dominant gravitational wave frequency peak in the postmerger phase to the radius of nonrotating neutron stars of a certain mass. We clarify the physical origin of secondary peaks and discuss a spectral classification scheme, based on their relative strength. Observational prospects for the dominant and the secondary peaks are also discussed. The threshold mass to black-hole collapse is connected by another empirical relation to the maximum mass and compactness of nonrotating neutron stars, which can be derived semi-analytically. The observation of GW170817 then sets an absolute minimum radius for neutron stars of typical masses, based only on a minimal number of assumptions.  

\end{abstract}

\section{Introduction}
\label{intro}

The very first detection of gravitational waves (GWs) from colliding neutron stars (NSs) in August 2017 represents the most recent highlight in the field of compact objects \cite{Abbott2017}. The Advanced LIGO-Advanced Virgo network of GW instruments observed a compact binary merger with a total mass of 2.74~$M_\odot$ at a distance of about 40~Mpc. Gamma rays were detected 1.7 seconds after the merger \cite{Abbott2017b,Abbott2017f,Goldstein2017,Savchenko2017}. Follow-up observations with optical telescopes identified an electromagnetic counterpart about 12 hours after the GW event close to the galaxy NGC 4993 \cite{Cowperthwaite2017,Kasen2017,Nicholl2017,Chornock2017,Drout2017,Smartt2017,Kasliwal2017,Kilpatrick2017,Perego2017,Tanvir2017,Tanaka2017}. This transient is called AT2017gfo. Finally, X-rays and radio emission were detected several days after the merger \cite{Troja2017,Margutti2017,Hallinan2017}. The GW measurement and the detection of accompanying electromagnetic emission have finally established the link between theoretical work on NS mergers and observations. This immediately led to some important insights.

For instance, the detection allows an estimate of the merger rate in the local Universe \cite{Abbott2017}. Employing the redshift of the host galaxy and the measured luminosity distance leads to an independent estimate of the Hubble constant, which is found to be compatible with other measurements \cite{Abbott2017}. 
The measured GWs in combination with detailed modeling of the signal revealed the total binary mass, constraints on the binary mass ratio and an upper limit on the tidal deformability of NSs \cite{Abbott2017,Abbott2019,PhysRevLett.121.161101,Chatziioannou2018,De2018,Carney2018}. The tidal deformability describes a ``matter effect'' on the GW signal and as such constrains uncertain properties of high-density matter, i.e.\ the equation of state (EoS) of NS matter. The measurement implies that nuclear matter cannot be very stiff and that NSs with masses of about 1.35~$M_\odot$ cannot be larger than about 14~km, e.g. \cite{Fattoyev2018,Raithel2018,TheLIGOScientificCollaboration2018}.

The exact interpretation of detected gamma rays, X-rays and radio emission in connection with GW170817 is not yet fully clear \cite{Abbott2017b,Abbott2017f,Troja2017,Margutti2017,Goldstein2017,Kasliwal2017,Hallinan2017,Bromberg2018,Gottlieb2018,Pozanenko2018,Lazzati2018,Granot2018,Lamb2018,Xie2018,Mooley2018,Ruan2018,Margutti2018,Nakar2018}. Long-term observations provide further insights. While the detected gamma-rays provide very strong support that gamma-ray bursts are connected with NS mergers~\cite{Paczynski1986,Eichler1989,Nakar2007,Berger2011}, the emission was subluminous compared to other gamma-ray bursts considering the distance of the event.

A wealth of observational data on NS mergers is now available and more data may become available soon. However, simulations and theoretical studies of the merger process are still critical, because the interpretation of these observations relies on these theoretical models. This includes the analysis of GW detections, which links the detailed evolution of the signal to physical properties of the merger, such as the individual binary masses and the EoS of NS matter.  Clearly, the importance of simulations will increase,  as more detailed observational data become available especially at high frequencies.

The properties of high-density matter, in particular the EoS around and beyond nuclear saturation density, are only incompletely known, because of the challenges to address the nuclear many-body problem and the corresponding nuclear interactions, e.g.~\cite{Lattimer2016,Oertel2017,Baym2018,Lalit2019,Tews2019}. Moreover, it is not clear whether additional particles, e.g.\ hyperons, occur at higher densities or whether even a phase transition to deconfined quark matter takes place in NSs. As a consequence of these theoretical uncertainties, the stellar properties of NSs are not precisely known. The stellar structure of nonrotating NSs is uniquely determined by the EoS through the Tolman-Oppenheimer-Volkoff equations, which describe  
relativistic hydrostatic equilibrium \cite{Tolman1939,Oppenheimer1939}. In particular, the stellar radius for a given mass, i.e.\ the mass-radius relation, the tidal deformability\footnote{The tidal deformability is defined by $\Lambda=\frac{2}{3}k_2 \left( \frac{c^2R}{G M} \right)^5$ with the tidal Love number $k_2$, the stellar radius $R$ and mass $M$~\cite{Hinderer2008,Hinderer2010}. $c$ and $G$ are the speed of light and the gravitational constant.} and the maximum mass of nonrotating NSs are not precisely known.  Beside recent studies in the context of GW170817, observational efforts to determine stellar properties of NSs are also not fully conclusive because they typically suffer from measurement errors and systematic uncertainties \cite{Lattimer2016,Oezel2016,Oertel2017}. 

Here, we will summarize our recent work that aims at understanding 
the EoS dependence of the GW signal by studying the impact of a sample of representative candidate EoSs on the merger dynamics and on the emission of gravitational radiation. By establishing a link between observable features of the GW signal and characteristics of the EoS, a GW detection can be employed to infer or at least constrain properties of high-density matter. Taking advantage of the correspondence between the EoS and stellar properties, it is convenient to use stellar parameters of NSs to characterize a given EoS. We thus directly link the GW signal with stellar properties of NSs.

This effort is not only relevant for a deeper understanding of fundamental properties of matter, but also for the comprehension of astrophysical processes and events which are influenced by the NS EoS such as core-collapse supernovae and NS cooling \cite{Yakovlev2004,Janka2012,Oertel2017}.

Many previous and current studies consider the impact of the EoS on the pre-merger phase, i.e.\ on the orbital dynamics before merging, the so-called ``inspiral'' \cite{Faber2002,Flanagan2008,Hinderer2010,Read2009,Damour2010,Damour2012,Bernuzzi2012,Read2013,DelPozzo2013,Wade2014,Agathos2015,Chatziioannou2015,Hotokezaka2016,Abbott2017,Chatziioannou2018,Abbott2019,PhysRevLett.121.161101,De2018,Carney2018}. In our studies we follow a different strategy and focus on the postmerger phase to devise methods which are complementary to the existing approaches. We stress the importance of developing alternative techniques to measure EoS properties. Although GWs track the bulk motion of matter and thus represent a particularly robust messenger of the EoS influence on the dynamics, the results of GW inference approaches that rely on current inspiral waveform models are not free of systematic uncertainties~\cite{Wade2014}. Difficulties arise in designing waveform models which are sufficiently accurate and reliable over a wide range of source parameters. 
An alternative method for EoS constraints is thus desirable to independently corroborate the interpretation of observational data,  while we remark that inference from the postmerger stage relies to a large extent on numerical modeling, which faces challenges as well, e.g. regarding the exact damping behavior and the inclusion of all relevant physical effects.  Moreover, the methods laid out in this review are complementary in the sense that they may probe regimes of the EoS which are not accessible during the pre-merger phase. The density increases during and after merging, and thus the postmerger GW signal may also reveal properties of the EoS at higher densities (see~\cite{Bauswein2019}). 

We note that the goal of this review article is to summarize recent personal work, without the intention of represeting a complete review of the field. The interested reader is referred to \cite{Faber2012,Bauswein2016,Baiotti2017,Metzger2017a,Paschalidis2017,2018arXiv180704480V,2018IJMPD..2743018F,Duez2019,Cowan2019} and references therein for additional reviews.
 
 {\it Outline:} We provide here an extended outline to help the reader to navigate through the sections. In Sect.~\ref{sec:stages} we summarize the dynamics and the outcome of NS mergers and some relevant physical effects. Sect.~\ref{sec:tool} provides some background information on the simulations which are discussed here. 

We describe a generic GW postmerger spectrum in Sect.~\ref{dominant} and point out the most prominent feature of the postmerger stage. This is a peak in the kHz range, which corresponds to the dominant oscillation mode of the postmerger remnant and promises the highest chance of detection among the postmerger features. Generally, postmerger GW emission is more challenging to detect compared to the gravitational radiation during the inspiral phase, which is dominanted by the orbital motion. Sect.~\ref{sec:fpeakrad} stresses that the freqency of the main postmerger GW peak is found empirically to correlate tightly with radii of non-rotating NSs for fixed individual binary masses. We also discuss that the frequency of the main feature increases with the total binary mass, while the binary mass has a much weaker influence on the frequency of the dominant peak. In turn, these dependencies offer the opportunity to determine NS radii if the total binary mass and the dominant postmerger GW frequency are measured. This is described in Sect.~\ref{sec:radcon}, where we also remark that a strong first-order phase transition may lead to a characteristic increase of the postmerger frequency. 

Sect.~\ref{sec:gwana} summarizes efforts to develop dedicated GW data analysis methods to extract the dominant postmerger GW frequency from a measurement. This will require instruments with better sensitivity in the kHz regime and may be achieved with the current instruments at design sensitivity or with projected upgrades to the existing detectors. Generally, the extraction of the main frequency with a precision of a few 10~Hz seems feasible for sufficiently near-by events. In Sect.~\ref{sec:morecons} we discuss that several of such measurements could provide additional information on the properties of high-density matter, e.g. the maximum mass of nonrotating NSs or the maximum density inside NSs. 

Sect.~\ref{secondary} describes the physical origin of the main peak and secondary features in the postmerger GW spectrum. There is good evidence that the dominant peak is generated by the quadrupolar fluid mode, while secondary features are created by a coupling between the quadrupolar and the quasi-radial oscillation mode of the remnant. Another secondary feature is dynamically produced by two massive outer bulges which orbit on the surface of the remnant for a few milliseconds after merging. The different mechanisms for secondary GW peaks can be differently strong leading to more pronounced or supressed features in the spectrum and other characteristics of the remnant evolution like for instance the maximum density or the minimum lapse function as approximate measures of the gravitational potential. Sect.~\ref{sec:class} combines these observations in a unified classification scheme of the postmerger GW emission and dynamics. Three different types of evolution can be distinguished depending on the presence or absence of secondary GW features. The occurrence of the different classes shows a clear dependence on the total binary mass and on the properties of the high-density EoS. Sect.~\ref{sec:seceos} presents how the frequencies of the secondary GW peaks depend on the EoS and the total binary mass. For a fixed total binary mass the secondary GW frequencies follow closely the main peak in the postmerger GW spectrum. 

In Sect.~\ref{collapse} we then emphasize that the stability of the merger remnant directly after merging depends sensitively on the total binary mass and the EoS. Importantly, the different outcomes of a binary coalescence, i.e. direct black-hole (BH) formation or a NS remnant, lead to different observational features allowing the distinction of the two cases for instance through the characteristics of the postmerger GW emission or the properties of the electromagnetic counterpart. Since the total binary mass is measurable form the GW inspiral phase, this motivates to introduce a threshold binary mass for prompt BH formation, which we discuss in Sect.~\ref{sec:thres}. By a number of GW detections the threshold mass may be measured or at least constrained in the future. Simulations show that the threshold binary mass depends with good accuracy on the maximum mass and radii of nonrotating NSs. This implies that a measurement of the threshold mass determines the maximum mass of nonrotating NSs if other observations measure NS radii with sufficient precision. In Sect.~\ref{sec:rns} we discuss that a semi-analytic model confirms the particular EoS dependence of the threshold mass. 

We describe in Sect.~\ref{sec:gw17} that a multi-messenger interpretation of GW170817 implies a lower limit on NS radii of about 10.7~km. This constraint employs the reasonable assumption that the merger did not result in the direct formation of a BH, which may be suggested by the electromagnetic emission. This conclusion is a direct consequence of the dependence of the threshold mass for direct BH formation on the maximum mass and radii of nonrotating NSs. Neutron-star radii must be larger than the limit because otherwise the remnant would have directly collapsed to a BH independent of the maximum mass. 

We conclude in Sect.~\ref{sec:sum} highlighting the complementarity of information about the high-density EoS, which can be obtained from the consideration of the postmerger evolution of NS mergers compared to other methods. The prospects to gain additional insight into the properties of high-density matter stresses the importance of GW detectors with increased sensitivity in the kHz range and GW data analysis models to infer features of the postmerger stage.
 
\section{Merger stages and dynamics}\label{sec:stages}
The phase preceding the merger is called ``inspiral'', where the orbital separation of the binary continuously shrinks as a result of the emission of GWs, which reduce the orbital energy and angular momentum. The inspiral proceeds increasingly faster because the GW emission becomes stronger with decreasing orbital separation. In this phase, the GW signal is essentially determined by the orbital dynamics resulting in a chirp-like signal with an increasing amplitude and an increasing frequency. Except for the very last phase, the inspiral can be well described by point-particle dynamics because the orbital separation is large compared to the stellar diameter. The inspiral time $\tau$, i.e.\ the time until merging, depends on the individual binary masses and very sensitively on the initial orbital separation $a$ with $\tau \propto a^4$. For the known binary systems, inspiral times between ${{\sim}}100$~Myrs and more than the Hubble time are found \cite{Lorimer2008}.

A NS binary reaches a GW frequency of ${{\sim}}10$~Hz only a few 10~seconds before merging, i.e.\ only then the system enters the sensitivity window of ground-based GW instruments, which ranges from a few 10~Hz to a few kHz (see, e.g.~\cite{Abbott2019}). The GW signal during the inspiral is dominated by the so-called chirp mass ${\mathcal{M}}_\mathrm{chirp}=\left(M_1 M_2 \right)^{3/5}/\left(M_1+M_2 \right)^{1/5}$ with $M_1$ and $M_2$ being the masses of the individual binary components. The mass ratio $q=M_1/M_2$ with $M_1\leq M_2$ enters the description of the dynamics and the waveform only at higher post-Newtonian order \cite{Creighton2011,Blanchet2014}. Therefore, it has a weaker impact on the signal, and mass-ratio effects become more pronounced in the last phase of the inspiral. Similarly, finite-size effects influence the signal only during the very last orbits before merging \cite{Damour2010,Blanchet2014}.

Since the chirp mass dominates the GW signal, ${\mathcal{M}}_\mathrm{chirp}$ is the parameter which is measured with the highest accuracy in comparison to the mass ratio and the individual masses of the binary components. See \cite{Abbott2016a,Abbott2016,Abbott2017e,Abbott2017d,Abbott2017c,Abbott2017} for examples. Measuring the mass ratio is crucial to determine the physical masses of the system $M_1$ and $M_2$. The chirp mass alone may provide only a estimate of the total mass  if the mass ratio is not well constrained (see e.g.\ Fig.~1 in \cite{Bauswein2016}).  In the case of GW170817 the total binary was found to be $M_\mathrm{tot}=2.74^{+0.04}_{-0.01}~M_\odot$, while the mass ratio $q=M_1/M_2$ was between 0.7 and 1~\cite{Abbott2017}. One should bear in mind that any GW detection provides the chirp mass with very good precision, while only for nearby mergers with high signal-to-noise (SNR) ratio the individual masses of the binary system can be determined. We will assume that the individual masses can be determined with the required accuracy, which is a reasonable assumption for systems where postmerger GW emission will become detectable.  As discussed in~\cite{Torres-Rivas2019}, an improvement in sensitivity by a factor of a few times is needed to detect postmerger emission at a few tens of Mpc. If the accuracy of mass measurements scales roughly with $(\mathrm{SNR})^{-1}$ or $(\mathrm{SNR})^{-1/2}$, individual binary component masses at such distances will be measured with a precision of per cent~\cite{Rodriguez2014,Farr2016,Abbott2017}. We note already here that for most aspects discussed below (e.g. the dominant postmerger GW frequency or the collapse behavior), the total binary mass has a much stronger impact than the mass ratio. Already with the current sensitivity the total binary mass of events at a few ten Mpc can be obtained with an accuracy of the order of one per cent~\cite{Rodriguez2014,Farr2016,Abbott2017}.

The orbital period prior to merging decreases to about 1~millisecond (the precise value depends on the individual masses and on the EoS of NS matter), and the stars exhibit strong tidal deformations. Because of the high orbital angular momentum, the stars coalesce with a relatively large impact parameter. The outcome of the merging depends critically on the total binary mass and the EoS. For relatively high masses the remnant cannot be stabilized against gravity and collapses to a BH on a time scale of less than ${\sim}1$~millisecond. This ``prompt collapse'' scenario  
 differs from the formation of a NS merger remnant, which occurs for lower total binary masses. The threshold between the direct formation of a BH and the formation of a NS remnant depends on the properties of NS matter, i.e.\ on the incompletely known EoS of high-density matter.

In the case of a prompt collapse a certain fraction of matter may become gravitationally unbound from the system, and a torus surrounding the central BH may form. If the merging results in a NS remnant, the central object initially consists of a rotating, highly deformed structure that is heavily oscillating. Since the remnant is rapidly rotating, the object can be stabilized against the gravitational collapse even if its total mass exceeds the maximum mass of nonrotating or uniformly rotating NSs. Since temperatures rise to a few 10 MeV, thermal pressure may contribute to the stabilization of the remnant. In particular, during the first milliseconds after merging, matter is ejected from the system, and a dilute halo and torus form around the central object. Generally, the system will evolve towards a state of uniform rotation, zero temperature and axisymmetry. Angular momentum redistribution and losses by GWs, mass ejection and neutrino cooling may result in a ``delayed collapse'' of the remnant. The life time of the remnant can be as low as a few milliseconds and depends sensitively on the total mass. For low total binary masses the product of the merger may be a massive rigidly rotating NS, which, however, may collapse as a result of magnetic spin-down on time scales of many seconds to minutes. For binaries with very low masses the final object may be stable. In all cases mass ejection continues at a lower rate by secular processes (neutrino-driven, viscously driven, magnetically driven) either from a BH torus forming after a prompt or delayed collapse or from the massive NS remnant. 
General reviews on NS mergers can be found for instance in \cite{Faber2012,Bauswein2016,Baiotti2017,Metzger2017a,Paschalidis2017,2018IJMPD..2743018F,Duez2019,Cowan2019}

\section{Simulation tool}\label{sec:tool}

While the inspiral phase can be modeled by a post-Newtonian expansion or by an effective one-body approach, the dynamical merging phase and the evolution of the postmerger remnant can only be adequately described by hydrodynamical simulations. This concerns in particular the merger outcome (prompt collapse, delayed collapse or no collapse), the GW emission of the postmerger phase and the mass ejection. The modeling of NS mergers requires a general-relativistic treatment because NSs are compact objects associated with a strong curvature of space-time.  
 Black-hole formation is an intrinsically general-relativistic process, and velocities during merging can reach a substantial fraction of the speed of light.

We performed simulations with a general-relativistic smooth particle hydrodynamics code (SPH)\cite{Oechslin2002,Oechslin2007,Bauswein2010,Bauswein2010a,Bauswein2010b,Just2015}. Within this approach the fluid is modeled by a set of particles of constant rest mass.  
The particles are advected with the flow and the hydrodynamical quantities are evolved on the position of the particles, i.e.\ comoving with the fluid. This Lagrangian formulation of hydrodynamics is particularly suitable for highly advective problems like NS mergers. It has the advantage of focusing computational resources on the most relevant parts of the fluid, instead of evolving large domains of an artificial atmosphere between and around the stars as in grid-based approaches to hydrodynamics.

The hydrodynamics scheme is coupled with a solver for the Einstein equations, which have to be evolved simultaneously to obtain the dynamical space-time metric. In the current implementation the metric is computed by employing the so-called conformal flatness condition, which imposes a conformally flat spatial metric \cite{Isenberg1980,Wilson1996}. This results in an approximate solution of the Einstein equations, but allows a very efficient computation of the self-gravity of the fluid. Assuming conformal flatness of the spatial metric however implies to explicitly neglect GWs, which are the driver of the inspiral. Therefore, the code incorporates a post-Newtonian GW backreaction scheme that computes corrections to the conformally flat metric and thus effectively mimics the losses of angular momentum and energy by GWs (see \cite{Oechslin2007} for details). The GW signal is extracted from the simulations by means of a modified version of the quadrupole formula which takes into account post-Newtonian corrections \cite{Oechslin2007,Blanchet1990}. Comparisons to codes computing the full solution of the Einstein equations show a very good agreement, in particular considering GW frequencies. {/bf For instance, the frequencies peaks of the postmerger GW spectra agree within a few per cent in} \cite{Bauswein2012a,Hotokezaka2013a,Takami2015,Maione2017}. The agreement may not be surprising, given that the conformal flatness condition yields exact results for spherical symmetry. 
The code also allows to simulate stellar matter in the presence of a BH. To this end a static puncture approach \cite{Brandt1997} is implemented \cite{Bauswein2010b,Just2015}. Despite the name, this scheme allows the BH to move and to conserve the total momentum.

To close the system of hydrodynamical evolution equations an EoS has to be provided, which describes the state of matter. Temperatures during merging can reach several ten MeV, and thermodynamical quantities of NS matter depend also on the composition, i.e.\ the electron fraction. These thermal and compositional effects have a substantial impact on the thermodynamical properties in NS mergers. Therefore, the EoSs used in this work describe the pressure and the energy density as function of the rest-mass density, temperature and electron fraction. Since NS matter involves complex microphysical models, the EoSs are provided as tables. These models are available in the literature (see e.g.\ \cite{Bauswein2012a,Bauswein2014a} for a compilation of different models employed in the simulations). Within this work, detailed compositional changes by weak interactions are not taken into account. Instead the initial electron fraction is advected, which may represent a reasonable approximation, since the impact of compositional changes on the bulk dynamics and the GW signal are relatively small. (See e.g. Figs.~12 and~13 in~\cite{Ardevol-Pulpillo2019} revealing that within a more elaborated treatment the electron fraction of the high-density material remains low, i.e. close to its initial value.) An accurate treatment of neutrino radiation effects is highly challenging and in any case requires certain approximations. An approximate treatment of weak interactions is implemented in the code~\cite{Ardevol-Pulpillo2019} but has not been applied in the studies presented here. 

Finally, the number of available EoS models which consistently provide the full dependence on temperature and electron fraction, is rather limited. A larger number of EoSs are given as barotropic relations describing NS matter at zero temperature and in neutrinoless beta-equilibrium. Those models can be employed in the code and are supplemented by an approximate treatment of thermal effects \cite{Bauswein2010}. This approach requires to specify a parameter $\Gamma_\mathrm{th}$, which regulates the strength of the thermal pressure contribution. By comparison to fully temperature-dependent EoS models a range for $\Gamma_\mathrm{th}$ can be fixed, and the impact of this choice on the final results can be quantified \cite{Bauswein2010}. Magnetic field effects are not included in the simulations but are likely to have only a negligible impact on the bulk dynamics and hence on the GW signal and dynamical mass ejection \cite{Giacomazzo2011,Kiuchi2014,Palenzuela2015,Kawamura2016}.

Typically, the simulations of NS mergers start a few revolutions before merging from a quasi-circular orbit. Within the standard NS binary formation scenarios circular orbits are expected because GW emission tends to reduce the eccentricity \cite{Camenzind2005,Lorimer2008}. Eccentric mergers may result from dynamical captures but are generally assumed to be less frequent \cite{Lee2010}. We do not consider eccentric binaries in this work. Initially, the stars are set up with zero temperature and in neutrinoless beta-equilibrium. Also, the intrinsic spin of the NSs is assumed to be zero because estimates have shown that tidal locking will not occur in NS binaries \cite{Bildsten1992,Kochanek1992}. Generally, the intrinsic spin of NSs is small compared to the orbital angular momentum, which justifies to assume an irrotational velocity field.  The fastest spinning NS in a NS binary system which will merge within the Hubble time, has spin period of 22~ms  (to be compared with the orbital period before merging of about 2~ms)~\cite{Burgay2003}. Because of magnetic dipole spin down the spin frequency will further decrease until the system merges. Another pulsar in a binary system has a spin period of 4~ms, but it is not clear whether the companion is a NS and the system will not coalesce within a Hubble time~\cite{Lynch2012}.

\section{Dominant postmerger gravitational-wave emission
}\label{dominant}
We will discuss NS merger simulations with a large, representative set of EoSs of NS matter. The main goal of such a survey is to devise procedures to infer unknown properties of NSs and of high-density matter from observables like the GW signal of a NS merger. The underlying idea is that the EoS affects the dynamics of a merger and therefore leaves an imprint on the GW signal. Whereas finite-size effects in the late inspiral phase have already been used to set EoS constraints based on GW170817~\cite{Abbott2017,Abbott2017a,Abbott2019,PhysRevLett.121.161101,De2018}, 
here we discuss a complementary approach, which is based on the GW signal of the postmerger phase.  In~\cite{Abbott2019} an unmodelled data analysis search has been performed to extract the postmerger GW emission. No signal was found, which is expected for the given distance of the event and the sensitivity of the instruments during the observations.

\begin{figure*}
\begin{center}
\resizebox{0.9\textwidth}{!}{
  \includegraphics{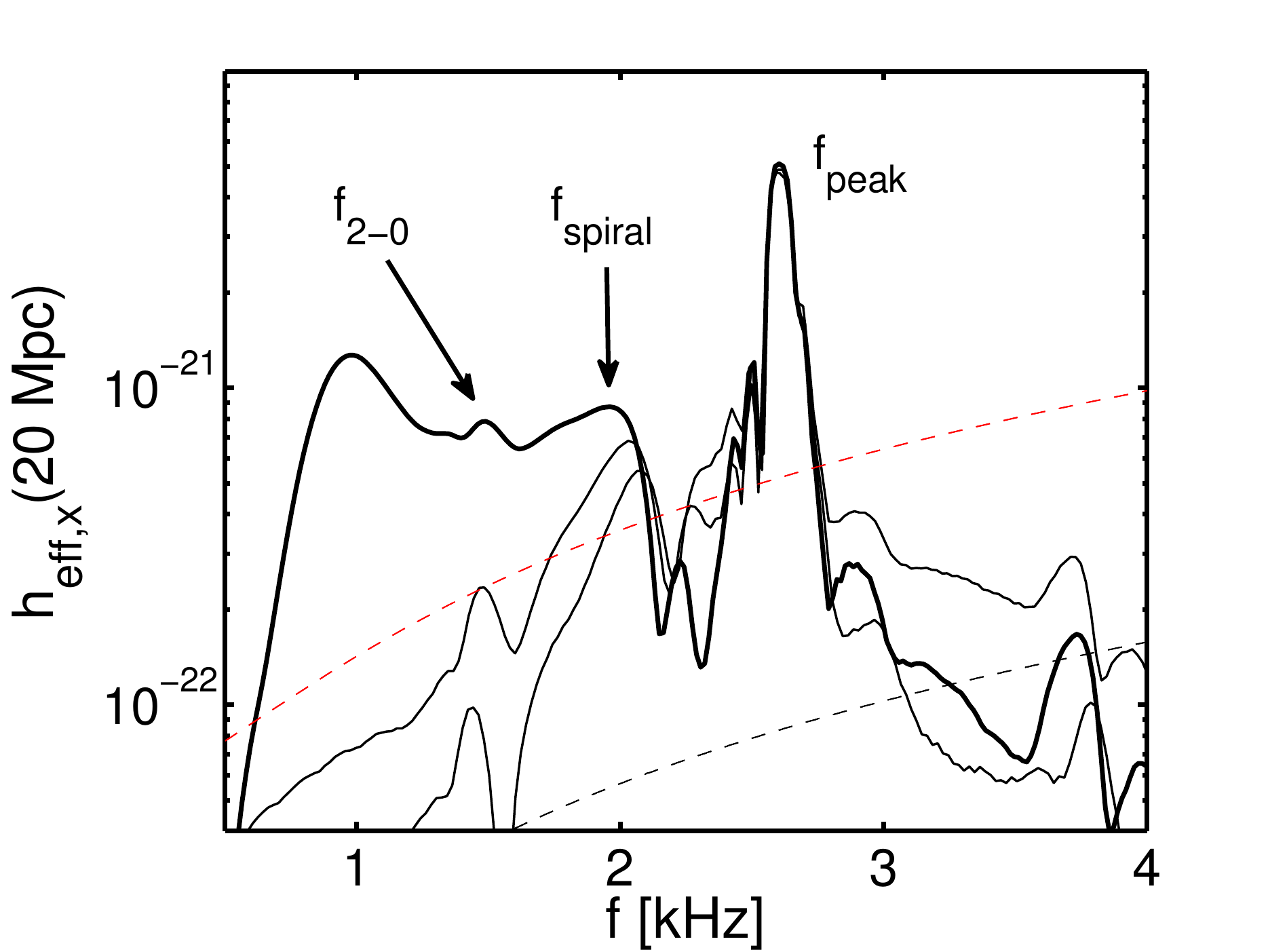}
}
\end{center}
  \caption{GW spectrum of the cross polarization of a 1.35-1.35~$M_\odot$ merger described by the DD2 EoS \cite{Hempel2010,Typel2010} along the polar direction at a distance of 20~Mpc.  $h_\mathrm{eff}=\tilde{h}\cdot f$ with $\tilde{h}$ being the Fourier transform of the dimensionless strain $h(t)$ and $f$ is frequency. The frequencies $f_\mathrm{peak}$, $f_\mathrm{spiral}$ and $f_{2-0}$ are distinct features of the postmerger phase, which can be associated with particular dynamical effects in the remnant. The thin solid lines display the GW spectra when the inspiral phase is ignored (with different cutoffs), revealing that the peaks are generated in the postmerger phase. Dashed lines show the expected design sensitivity curves of Advanced LIGO \cite{Harry2010} (red) and of the Einstein Telescope \cite{Hild2010} (black). Figure  from \cite{Bauswein2016}.}
  \label{fig:spectrum}
\end{figure*}

The most likely outcome of a NS merger is the formation of a meta-stable, differentially rotating NS remnant. A typical GW spectrum of such a case is shown in Fig.~\ref{fig:spectrum} for a NS merger of two stars with a mass of 1.35~$M_\odot$ each. The signal is extracted from a simulation with the DD2 EoS \cite{Hempel2010,Typel2010}. The low-frequency part of the spectrum is predominantly shaped by the inspiral phase, keeping in mind that the hydrodynamical simulations start only a few orbits before merging, which is why the shown spectrum significantly underestimates the power at lower frequencies. In this example, during the inspiral the GW frequency reaches about $\sim 1$ kHz when the amplitude becomes strongest and the binary system enters the merger phase. See for instance Figs.~5 and~6 in \cite{Bauswein2012a} for the GW amplitude in the time domain.

The postmerger spectrum exhibits several distinct peaks in the kHz range, which are connected to certain oscillation modes and dynamical features of the postmerger remnant. With an appropriate windowing of the signal, these peaks can be clearly associated with the postmerger stage. In terms of the effective GW amplitude $h_{\rm eff}=\tilde{h}(f)\cdot f$ (where $f$ is frequency and $\tilde h(f)$ the Fourier transform), there is a dominant oscillation frequency $f_\mathrm{peak}$, which is a generic and robust feature and which occurs in all merger simulations that do not result in a prompt formation of a BH.  The dominant postmerger peak is observationally the most relevant feature of the postmerger spectrum, since typically it has the highest signal to noise ratio of all distinct postmerger features. The secondary peaks ($f_\mathrm{spiral}$ and $f_{2-0}$) will be discussed in Sec. \ref{secondary}.

\section{Gravitational-wave frequency-radius relations}\label{sec:fpeakrad}
The properties of the dominant postmerger frequency peak and, in particular, its EoS dependence can only be assessed by means of hydrodynamical simulations. It was known that the frequency $f_\mathrm{peak}$ is affected by the EoS, which was concluded from calculations with a small number of candidate EoSs \cite{Shibata2005,Shibata2006,Shibata2005a,Oechslin2007a}. That $f_\mathrm{peak}$ depends in a {\it specific way} on the high-density EoS was shown in \cite{Bauswein2012,Bauswein2012a}, which in turn implies that $f_\mathrm{peak}$ can be employed to place EoS constraints. This was shown by an extensive set of merger simulations for a large, representative number of EoSs. This sample of candidate EoSs covers the full range of viable EoS models in terms of their resulting stellar properties (see Fig.~4 in \cite{Bauswein2012a} or Fig.~\ref{fig:radcon}), but excluding strong phase transitions (see \cite{Bauswein2019}).

As described in Sec.~\ref{intro}, the binary masses of NS mergers can be obtained from the inspiral phase. Especially for merger events which are sufficiently close to detect postmerger GW emission, the individual masses will be determined with good precision \cite{Rodriguez2014,Farr2016,Abbott2017}. One can thus consider the EoS dependence of $f_\mathrm{peak}$ for fixed individual binary mass configurations. We start by focusing on 1.35-1.35~$M_\odot$ mergers and discuss variations of the binary parameters afterwards. These systems are considered to be the most abundant in the binary population according to observations of NS binaries and theoretical models of the population (population synthesis) \cite{Lattimer2012,Dominik2012}. A total mass of 2.7~$M_\odot$ is also in line with the total mass of GW170817. Small deviations from mass symmetry do not lead to significant differences in the spectrum compared to the equal-mass case, while a mass ratio of 0.7 has only a moderate impact (see \cite{Hotokezaka2013a}).

\begin{figure*}
\begin{center}
\resizebox{1\textwidth}{!}{
\includegraphics{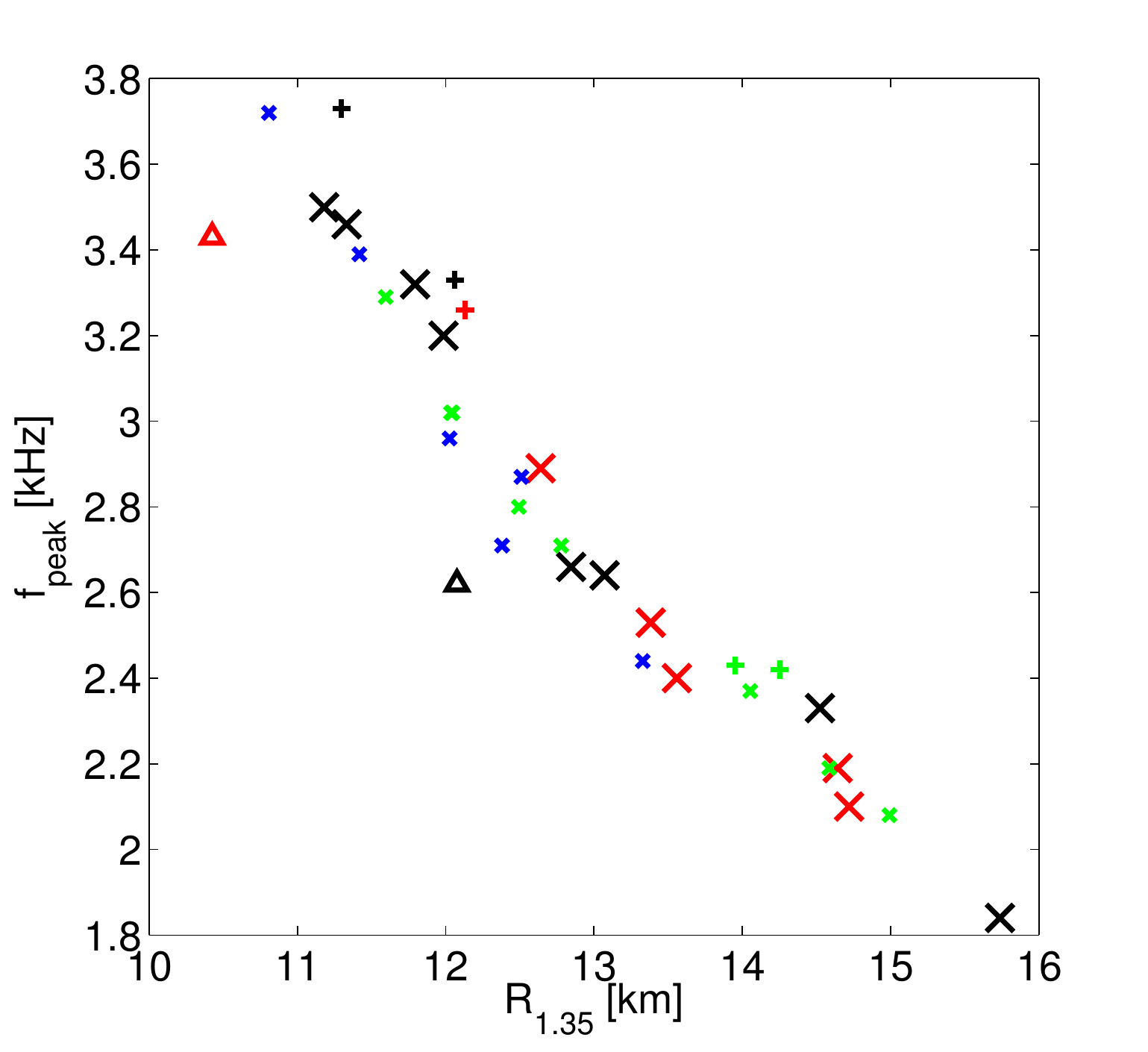}
\includegraphics{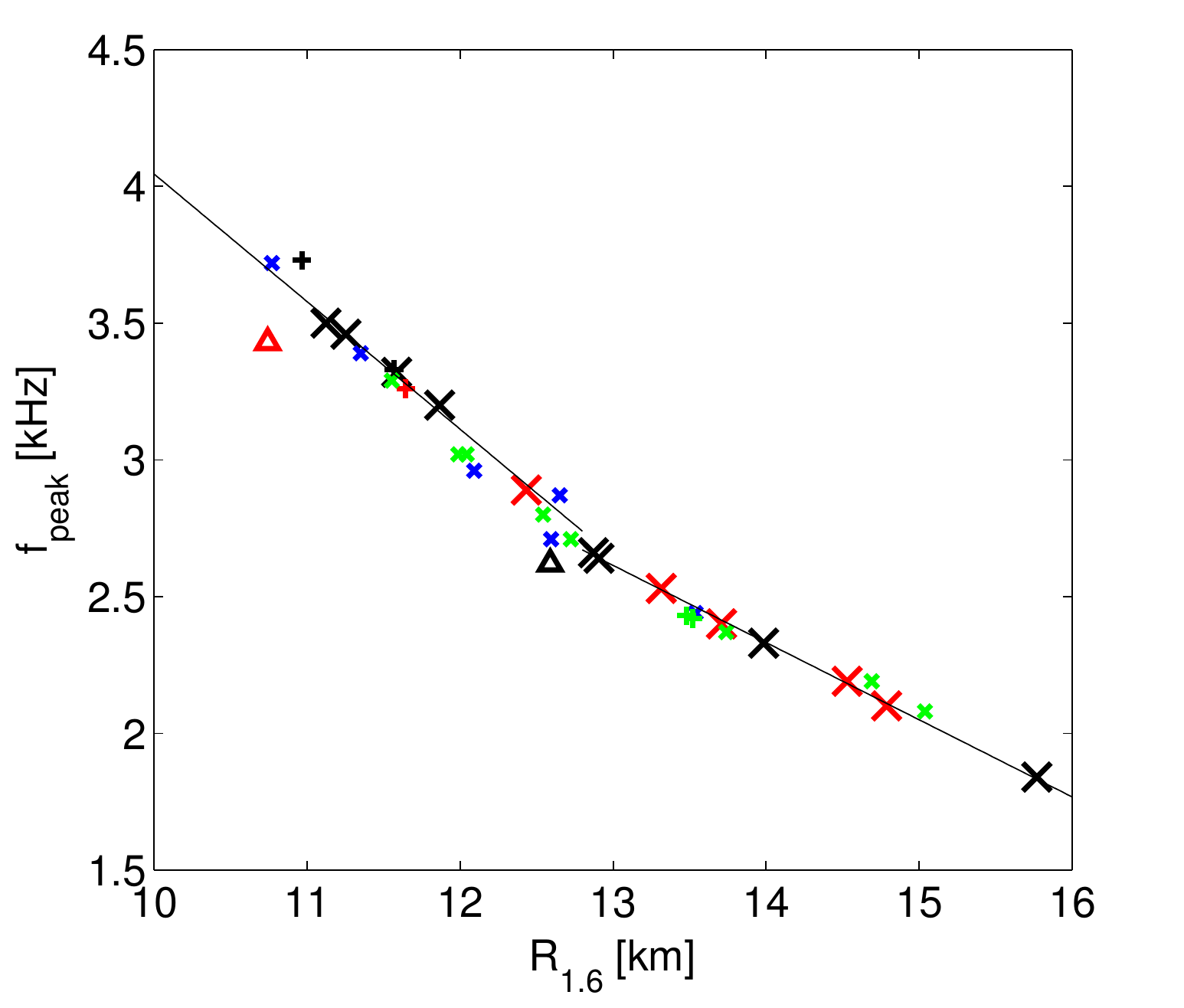}
}
\end{center}
  \caption{Peak frequency of the postmerger GW emission versus the radius of nonrotating NSs with a mass of 1.35~$M_\odot$ (left panel) and 1.6~$M_\odot$ (right panel) for different EoSs. These figures include only data from 1.35-1.35~$M_\odot$ mergers. Figures  from~\cite{Bauswein2012a}, which should be consulted for the detailed nomenclature. Triangles mark models of absolutely stable strange quark matter.}
  \label{fig:fpeakrad}
\end{figure*}

For a fixed binary mass configuration, the EoS dependence is determined by investigating empirical relations between the dominant postmerger frequency $f_\mathrm{peak}$ and EoS properties. Stellar parameters of nonrotating NSs are uniquely linked to the EoS through the Tolman-Oppenheimer-Volkoff equations, and properties such as NS radii represent integral properties of a given EoS. Such EoS characteristics turn out to be particularly suitable to describe the GW emission of NS mergers. Relating the peak frequency $f_\mathrm{peak}$ of 1.35-1.35~$M_\odot$ mergers with the radius $R_{1.35}$ of a nonrotating NS with 1.35~$M_\odot$ shows a clear correlation (see Fig.~\ref{fig:fpeakrad} and also Fig.~4 in \cite{Bauswein2012} and Fig.~12 in \cite{Bauswein2012a}). This relation specifically connects the radius of the inspiraling original stars with the oscillation frequency of the postmerger remnant, an object of roughly twice the mass of the individual stars. It is therefore plausible to explore relations between the peak frequency $f_\mathrm{peak}$ of 1.35-1.35~$M_\odot$ mergers and radii of nonrotating NSs with a higher fiducial mass (see Figs. 9 to 12 in \cite{Bauswein2012a}). All these relations exhibit a scaling between $f_\mathrm{peak}$ and the respective radius.

Figure~\ref{fig:fpeakrad} (right panel) shows the peak frequency as function of the radius $R_{1.6}$ of a nonrotating NS with 1.6~$M_\odot$ for the equal-mass mergers with a total mass of 2.7~$M_\odot$. The relation can be written as
\begin{equation}
f _ { \mathrm { peak } } = \left\{ \begin{array} { l l } 
           { - 0.2823 \cdot R _ { 1.6 } + 6.284, } & { \textrm { for } f _ { \rm peak  } < 2.8 \mathrm { kHz } }, \\ 
           { - 0.4667 \cdot R _ { 1.6 } + 8.713, } & { \textrm { for } f _ { \rm peak  } > 2.8 \mathrm { kHz } }. \end{array} \right.
\end{equation}
In \cite{Bauswein2012a} this relation is found to be the most accurate, when compared to relations that employ $R_{1.35}$, $R_{1.8}$ or the radius $R_\mathrm{max}$ of the maximum-mass configuration of nonrotating NSs. Here, the maximum deviation of the data points from a least-square fit is considered as figure of merit to assess the quality and accuracy of the relations. For $R_{1.6}$ the maximum scatter is less than 200~m. Note that some of these EoSs are already excluded by GW170817. Hence, an improved fit may be obtained by including current information on viable EoS models.

\begin{figure*}
\begin{center}
\resizebox{0.9\textwidth}{!}{
\includegraphics{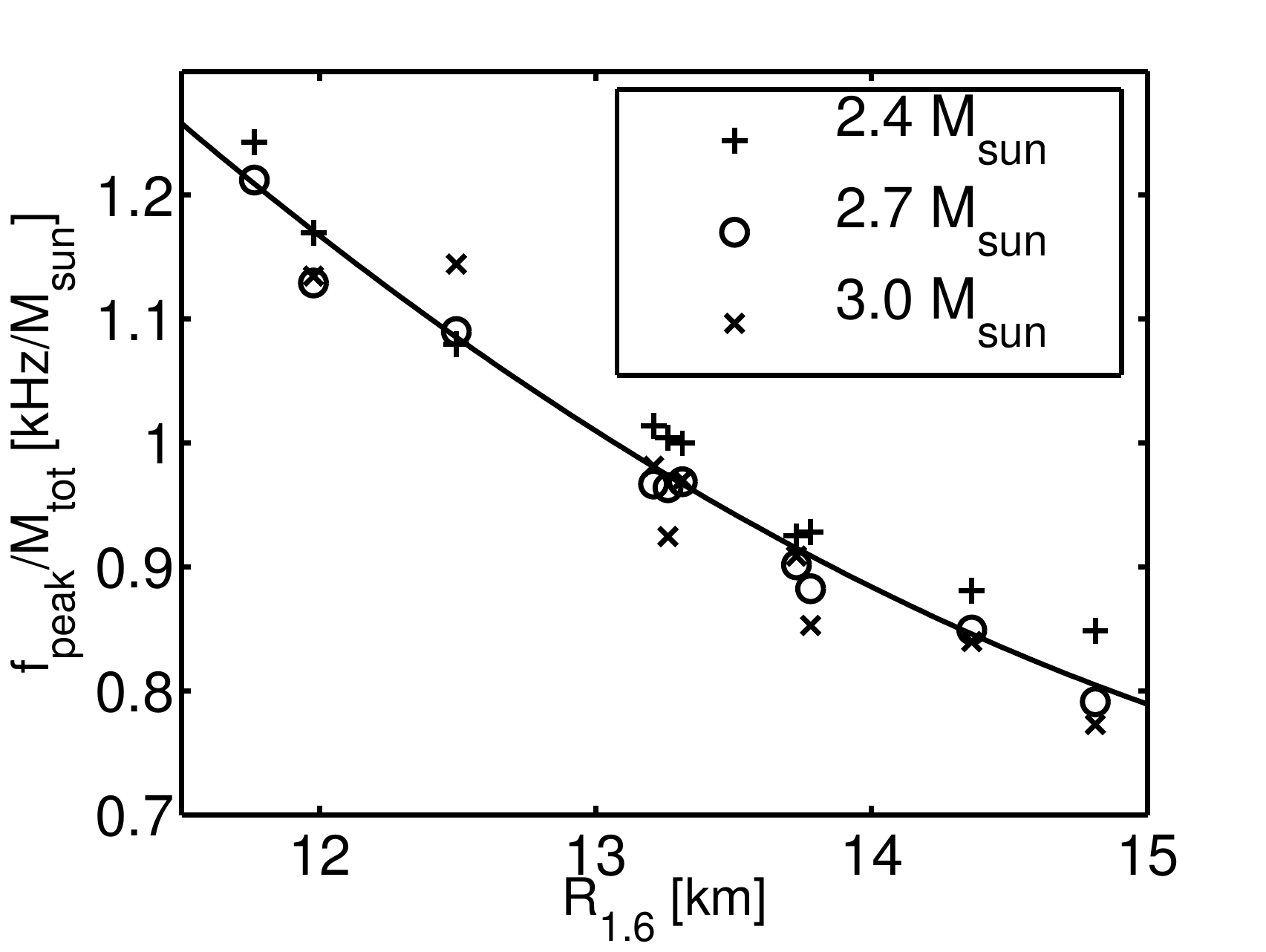}
}
\end{center}
  \caption{Rescaled dominant postmerger GW frequency $f_{\rm peak}/M_{\rm tot}$ as function of the radius $R_{1.6}$ of a nonrotating NS with a gravitational mass of $1.6M_\odot$ for different EoSs and different total binary mass and a mass ratio of unity. Figure  from~\cite{Bauswein2016}.}
  \label{fig:fpeakscaled}
\end{figure*}

Similar scalings between $f_\mathrm{peak}$ and NS radii exist also for other fixed individual binary masses, e.g.\ 1.2-1.2~$M_\odot$, 1.2-1.5~$M_\odot$ or 1.5-1.5~$M_\odot$ mergers and a single relation, scaled by the total mass is \cite{Bauswein2016}
\begin{equation}
f _ { \rm peak  } / M _ { \mathrm { tot } } = 0.0157 \cdot R _ { 1.6 } ^ { 2 } - 0.5495 \cdot R _ { 1.6 } + 5.5030,
\end{equation}
see Fig. \ref{fig:fpeakscaled} and Figs.~22 to~24 in \cite{Bauswein2012a}. See~\cite{Bernuzzi2015} for a similar rescaling but with the tidal coupling constant.
 
It is understandable that the radius of a NS somewhat more massive than the inspiraling NSs yields the tightest relation between $f_\mathrm{peak}$ and the radius. The reason is that the central density of the merger remnant is higher than the central density of the individual stars. Thus, the radius $R_{1.6}$ represents the EoS better within the density range of a merger with $M_\mathrm{tot}=2.7~M_\odot$ (see also discussion and figures in~\cite{Bauswein2019}). On the other hand, the evolution of the central density of the postmerger object, while typically strongly oscillating (see Fig.~15 in \cite{Bauswein2012a}), remains significantly below the maximum central density of nonrotating NSs (see also~\cite{Bauswein2019}), which is why the relation  between $f_\mathrm{peak}$ and $R_\mathrm{max}$ shows a relatively large scatter.

Higher $M_\mathrm{tot}$ result in higher peak frequencies. This is understandable because more massive remnants are more compact and thus oscillate at higher frequencies. One also recognizes that peak frequencies of high-mass mergers show tighter correlations with radii of relatively massive NSs, while the postmerger GW emission of mergers with lower total binary mass is well described by radii of nonrotating NSs with relatively small masses~\cite{Bauswein2012a}. In line with the arguments above, this observation is explained by the different density regimes which are probed by merger remnants of different total masses. 

Finally, the qualitative behavior of the frequency-radius relations is intuitive: softer EoSs, which lead to smaller radii of nonrotating NSs, also imply more compact merger remnants, which oscillate at higher frequencies (see Fig.~13 in \cite{Bauswein2012a}). As argued in Sec.~\ref{secondary}, the dominant oscillation is associated with the fundamental quadrupolar fluid mode. For nonrotating stars it is known that this mode scales with the mean density $\sqrt{M/R^3}$ \cite{Andersson1998}. This is why for fixed-mass sequences a strong radius dependence may be expected, keeping in mind that rapid rotation as in the merger remnant introduces significant corrections to the oscillation frequencies of nonrotating stars \cite{Doneva2013a}. Moreover, the mass of merger remnants typically exceeds the maximum mass of nonrotating NSs, and thus the oscillation frequencies of the remnant cannot be directly connected to oscillation modes of a nonrotating NS of the same mass. The corrections by rotation and the extrapolation to higher masses are, however, likely to depend in a continuous manner on the EoS, which may then explain the observed relations. A detailed investigation of oscillation modes of differentially rotating merger remnants is still to be developed.

\section{Radius measurements and EoS constraints}\label{sec:radcon}

The importance of these empirical correlations lies in the possibility to use them for a NS radius measurement when $f_\mathrm{peak}$ has been extracted from a GW observation. A measured peak frequency can be converted to a radius measurement by means of the frequency-radius relation. The maximum scatter in the relation should be taken into account as part of the systematic error of this measurement. A priori it is not clear how well the true EoS of NS matter follows the empirical correlation, which is built on basis of a set of viable candidate EoSs. However, if this sample of EoSs includes the most extreme models which are considered to be compatible with current knowledge, one may expect that the maximum deviation in the relation provides a conservative estimate of the systematic error of the frequency-radius inversion. In this context, it is worth mentioning 
that even absolutely stable strange quark stars \cite{Bodmer1971,Witten1984} follow the frequency-radius relations (triangles in Fig.~\ref{fig:fpeakrad}, see also Figs.~9 to~12 in \cite{Bauswein2012a}). We remark that the two models do consider bare strange stars and neglect a possible nuclear crust. In principle, strange stars may have a nuclear crust with densities up to the neutron-drip density, which would increase their radii by a few hundred meters. This crust would hardly affect the GW frequency because it contains only a small mass. Hence, the consideration of a nuclear crust would likely render the data for strange stars even more compatible with the $f_\mathrm{peak}$-radius relation of ordinary NSs. Given the significant qualitative differences between EoSs of absolutely stable quark matter and EoSs of hadronic NSs, the tight scaling between the dominant GW oscillation frequency and radii of nonrotating NSs represents a very robust correlation. 

\begin{figure*}
\begin{center}
\resizebox{0.9\textwidth}{!}{
\includegraphics{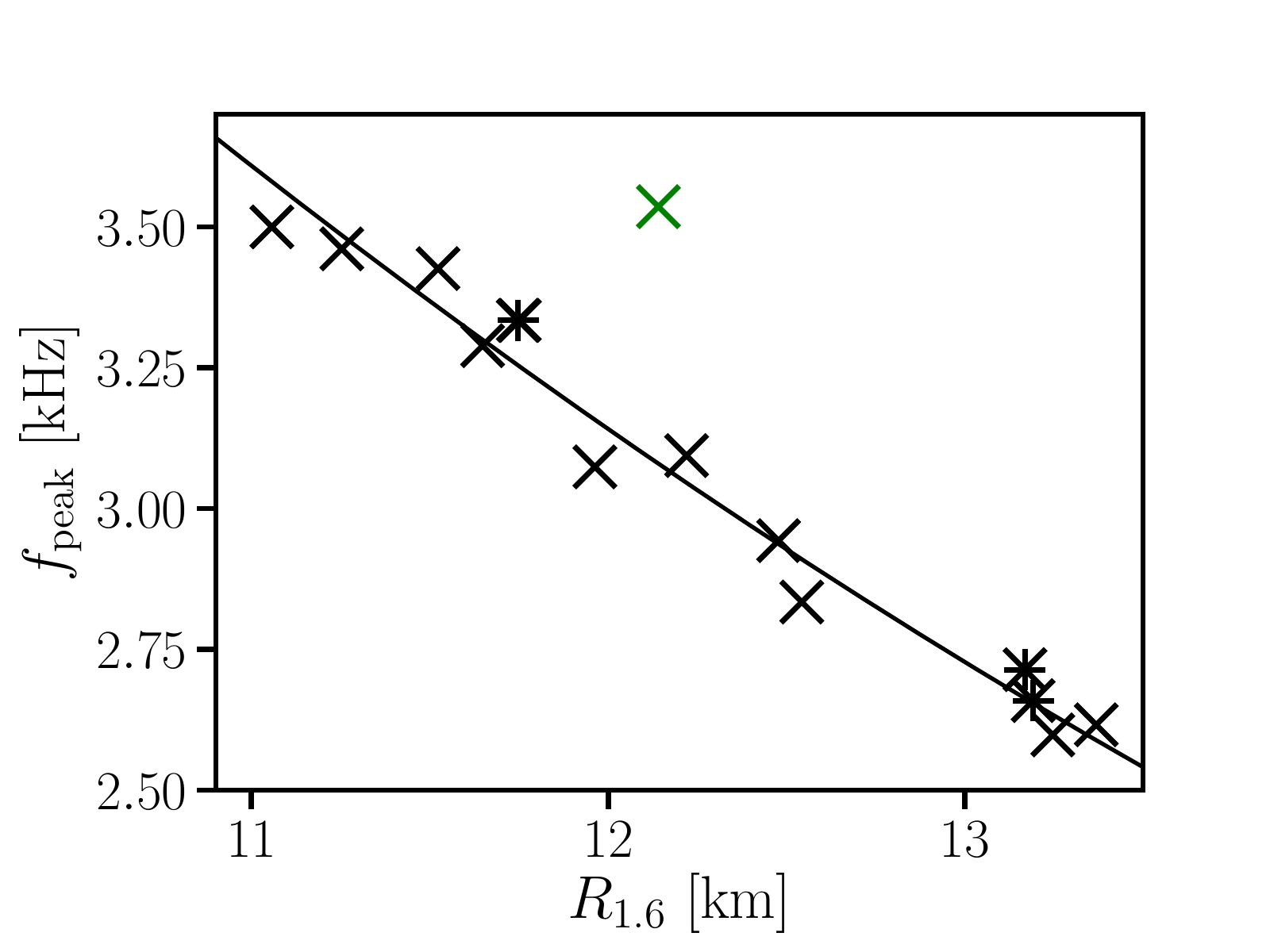}
}
\end{center}
  \caption{Dominant postmerger GW frequency $f_{\rm peak}$ as function of the radius $R_{1.6}$ of a nonrotating NS with $1.6 M_\odot$ for $1.35-1.35 M_\odot$ binaries. The DD2F-SF model (which exhibits a strong, first order phase transition to quark matter) is shown by a green symbol. Asterisks mark hyperonic EOSs. The solid curve provides a second-order polynomial least squares fit to the data (black symbols, excluding DD2F-SF). Models incompatible with GW170817 are not shown. Figure from~\cite{Bauswein2019}.}
  \label{fig:fpeakphase}
\end{figure*}

In \cite{Bauswein2019} an observable imprint of a first-order hadron-quark phase transition at supranuclear densities on the GW emission of NS mergers was identified. Specifically, the dominant postmerger GW frequency $f_{\rm peak}$ may exhibit a significant deviation from the empirical relation between $f_{\rm peak}$ and the radius $R_{1.6}$ (see Fig.~\ref{fig:fpeakphase}) if a strong first-order phase transition leads to the formation of a gravitationally stable, extended quark matter core in the postmerger remnant. A similar deviation exists if $f_{\rm peak}$ is considered as a function of the tidal deformability of $1.35 M_\odot$ NSs, see Fig. 3 in \cite{Bauswein2019}. Such a shift of the dominant postmerger GW frequency compared to the tidal deformability measured from the inspiral could be revealed by future GW observations, which would provide evidence for the existence of a strong first-order phase transition in the interior of NSs\footnote{In this case the empirical relation between $f_{\rm peak}$ and $R_{1.6}$ would provide a firm lower bound on NS radii.}. Note, however, that depending on the exact properties of the phase transition the impact on the merger dynamics and the GW signal might be significantly different, e.g.~\cite{Most2019}.

Similar to the frequency-radius relations, one can also explore the dependence of $f_\mathrm{peak}$ on other EoS characteristics. Examples are shown in \cite{Bauswein2012a} in Figs.~19 to 20 revealing an approximate scaling between $f_\mathrm{peak}$ and the maximum central energy of nonrotating NS or the speed of sound at 1.85 times the nuclear saturation density. The relation between the peak frequency and the pressure at 1.85 times the nuclear saturation density is particularly tight. Based on this finding the pressure at this fiducial density may be determined with a precision of about 10 per cent. This result may not be surprising given that NS radii are known to scale with the pressure at densities beyond saturation \cite{Lattimer2007}.

Asymmetric binaries lead to peak frequencies which are very similar to the dominant oscillation frequency of the equal-mass merger with the same total mass. This is not surprising because the oscillation frequency is determined by the stellar structure of the merger remnant, which is predominantly affected by the total mass and to a smaller extent by the mass ratio of the progenitor stars. Ref. \cite{Bauswein2016} shows a frequency-radius relation for a fixed chirp mass but varied mass ratio. The relatively tight correlation implies that a frequency-to-radius conversion is possible even if no information on the mass ratio is available but only an accurately measured chirp mass. 

Finally, the same study shows that the intrinsic spin of the progenitor stars has only a negligible impact on $f_\mathrm{peak}$ (see Fig.~15 in \cite{Bauswein2016}). The small influence of intrinsic rotation is somewhat in conflict with an apparently larger effect seen in \cite{Bernuzzi2014}, but is fully in line with other studies \cite{Kastaun2015,Dietrich2017}. Physically, a small impact of the initial spin makes sense given that the orbital angular momentum of the binary provides the majority of angular momentum of the remnant.

The existence of peak frequency-radius relations for other fixed binary masses and the relative insensitivity to the mass ratio or the intrinsic spin is an important finding for the actual application of these relations to radius measurements. As previously mentioned the total mass can be determined relatively well from the inspiral GW signal, while the mass ratio and the intrinsic spin are more difficult to measure. See, e.g., the extracted parameters for GW170817, for which $M_\mathrm{tot}=2.74^{+0.04}_{-0.01}~M_\odot$~\cite{Abbott2017}. After a GW detection, a peak frequency-radius relation has to be constructed based on simulation data for the corresponding binary parameters deduced from the inspiral phase.

\section{Gravitational-wave data analysis and detections}\label{sec:gwana}

The existence of the empirical relations described above has been confirmed and has triggered a lot of follow-up work by other groups, e.g.~\cite{Hotokezaka2013a,Takami2015,Bernuzzi2015,Palenzuela2015,Foucart2016,Maione2017,2016CQGra..33q5009M}. Moreover, it has motivated efforts to devise GW data analysis strategies for measuring $f_\mathrm{peak}$ \cite{Clark2014,Clark2016,Yang2018,Chatziioannou2017,Bose2018,Torres-Rivas2019}. A search for postmerger GW emission has been conducted for GW170817 employing also waveforms from the original studies presented above \cite{Abbott2017a}. No statistically relevant detection was reported, which, however, is expected considering the distance of GW170817 and the current sensitivity of the existing GW detectors. The instruments will reach their design sensitivity within the next years, which implies that a measurement of $f_\mathrm{peak}$ may be within reach if a GW event at a distance similar to the one of GW170817 is detected~\cite{Torres-Rivas2019}.

In \cite{Clark2014} an unmodeled burst search algorithm has been employed to recover our simulated waveforms which were injected in real recorded detector data containing only noise rescaled to the anticipated design sensitivity. This first study showed that a {\it morphology-independent} algorithm is able to measure the peak frequency $f_\mathrm{peak}$ with an accuracy of about 10~Hz at distances of ${\sim} 4 - 12$~Mpc with second-generation detectors. Thus, the statistical error in a radius measurement through $f_\mathrm{peak}$ should be expected to be small.

A more sensitive search can be devised by making certain assumptions about the signal morphology. For instance, in \cite{Clark2016} we consider a set of simulations to test the performance of a {\it principal component analysis} of candidate waveforms. Such an approach results in a detection horizon of 24 to 39~Mpc at design sensitivity of second-generation detectors, which is (given the GW170817 detection), within the regime where postmerger GW measurements become conceivable. 

Another method largely independent of assumptions about the signal morphology is based on a decomposition in wavelets \cite{Chatziioannou2017}. This method was updated in \cite{Torres-Rivas2019}, by including the available information from the pre-merger part of the signal. The main conclusion is that if a signal of similar strength to GW170817 is observed when the second-generation detectors have been improved by about 2 - 3 times over their design sensitivity in the kHz regime, then it will be possible to extract the dominant frequency component of the postmerger phase (with further improvements and next-generation detectors the subdominant frequencies will also be detectable). Thus, postmerger signals could be brought within our reach in the coming years, given planned detector upgrades, such as A+, Voyager, and the next-generation detectors~\cite{Punturo2010,Hild2011,Miller2015,Voyager2015,Martynov2019}.

Note that a first assessment of GW data analysis aspects of the postmerger phase has been presented in \cite{Bauswein2012a}, using a Fisher information matrix. While a Fisher information matrix approach is not fully applicable in the low SNR regime, the resulting accuracy and detectability inferred from this method are roughly consistent with the aforementioned more sophisticated methods. Further GW data analysis methods are continued to be developed, for example approaches that combine signals from several events to increase the overall sensitivity \cite{Yang2018,Bose2018} and approaches where a hierarchical model is trained on numerical-relativity simulations \cite{2018arXiv181111183E}.

The previous and ongoing work on GW data analysis methods marks the last component of a complete pipeline for the measurement of NS properties through postmerger GW emission. We remark that the total error of the measurement includes a statistical error from the $f_\mathrm{peak}$ determination and systematic errors. The latter include the maximum scatter in the frequency-radius relation for the measured individual binary mass and an uncertainty associated with the simulations, which rely on certain assumptions and approximations. A more precise determination of the different contributions is a task for the future. Overall, oscillation frequencies represent a rather robust bulk property of merger remnants, since they are essentially determined by the stellar structure. The total error of a radius measurement through the postmerger GW emission will be on the order of a few hundred meters \cite{Bauswein2012a,Clark2016}. The robustness and the complementarity of our method in comparison to inspiral methods relies on the fact that it does not require a detailed understanding of the phase evolution of the GW signal. Therefore, our method for NS radius measurements provides an interesting alternative to existing approaches based on the late inspiral phase. Finally, to quantify the prospects for a detection of $f_\mathrm{peak}$ the exact damping behavior of the postmerger GW signal has to be investigated in more detail.

\section{Further EoS constraints}\label{sec:morecons}
Measuring postmerger GW emission and determining the peak frequency clearly represent a challenge because a relatively loud signal is required. However, a single detection suffices to yield a robust radius measurement. Interestingly, even more stringent EoS constraints can be obtained if several detections of postmerger GWs are achieved, which may be possible with the current generation of ground-based GW instruments for a sufficiently long observing time.

\begin{figure*}
\begin{center}
\resizebox{1.0\textwidth}{!}{
\includegraphics{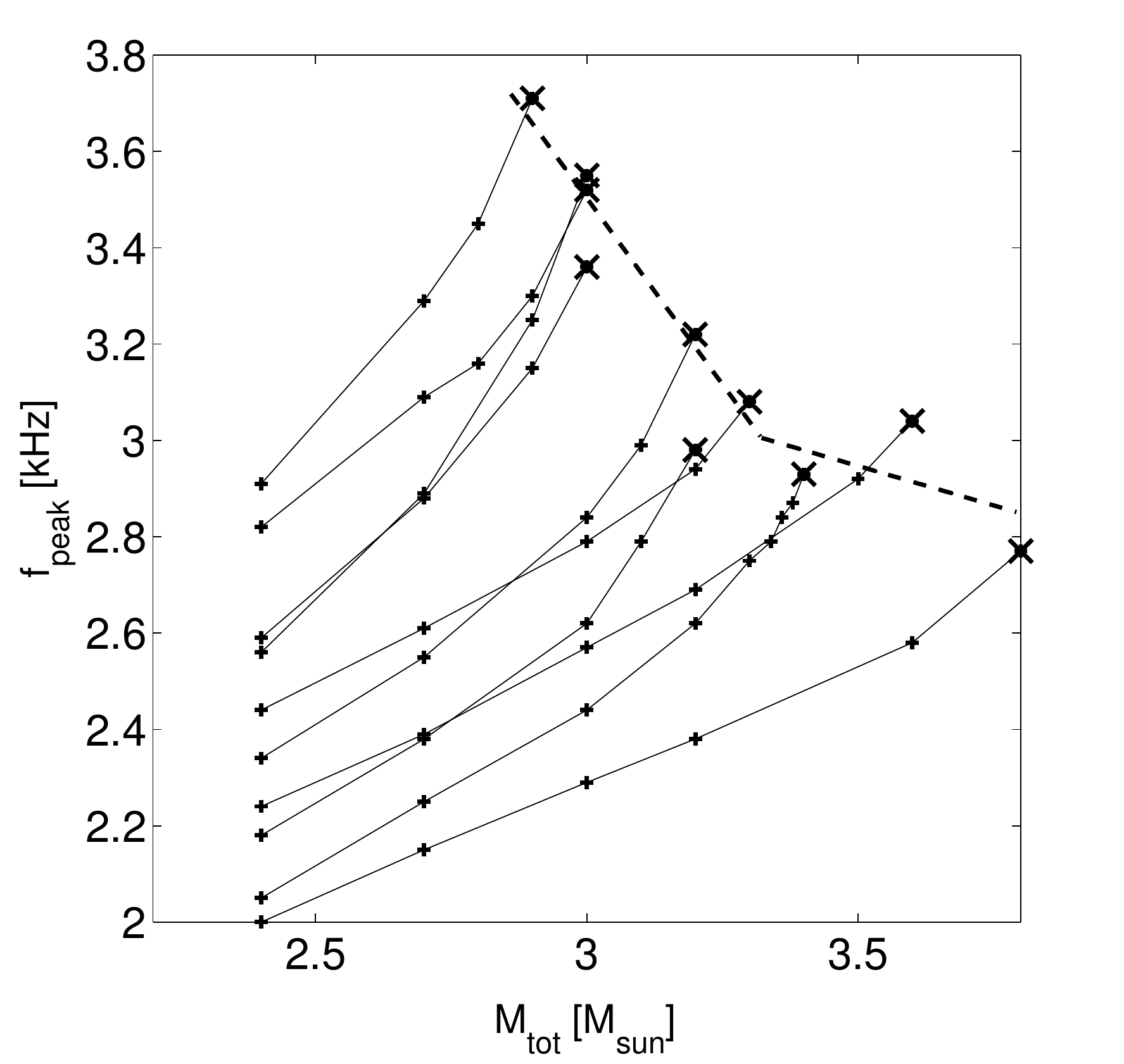}
\includegraphics{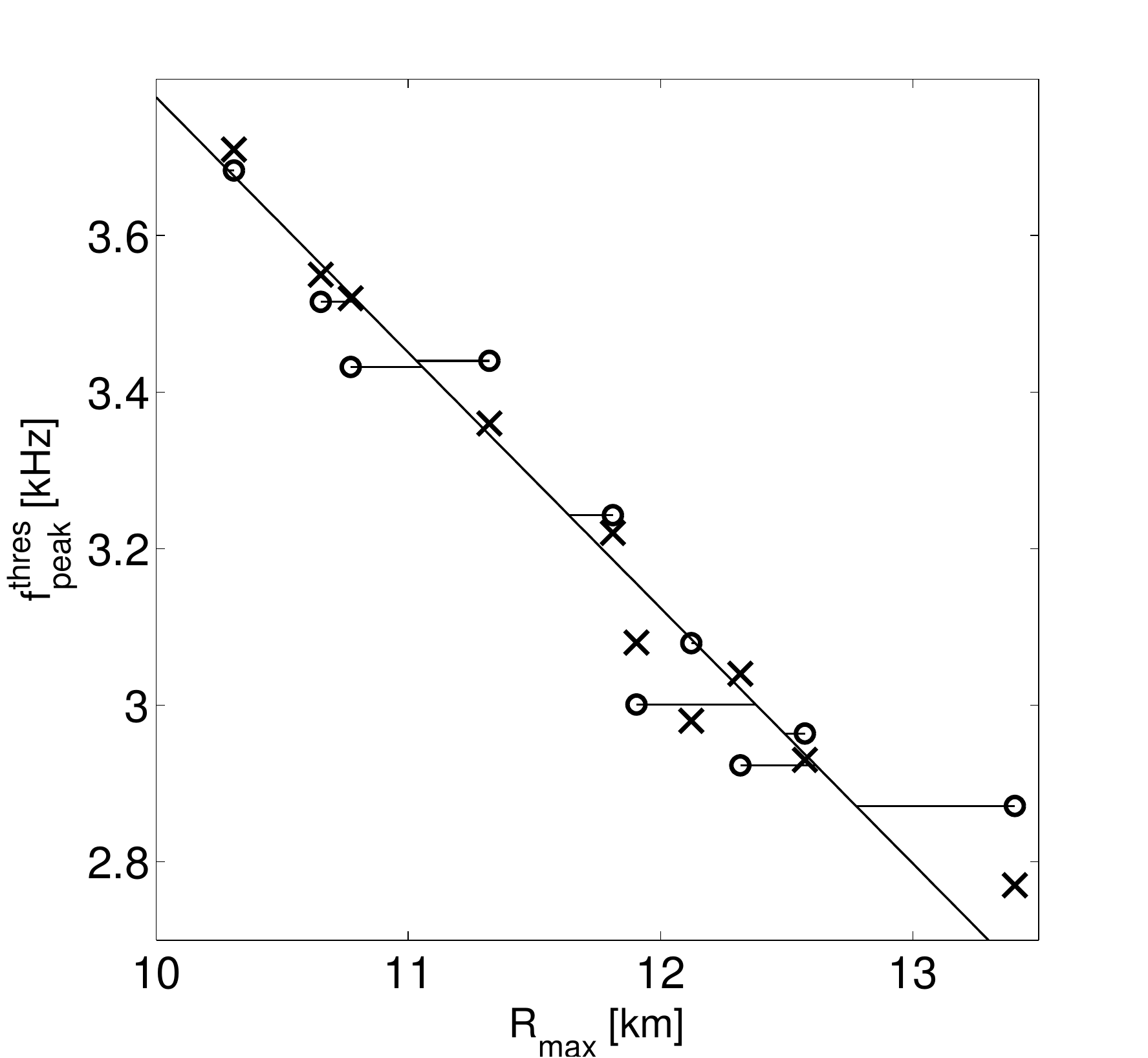}
}
\end{center}
  \caption{Left panel: Dominant postmerger GW frequency $f_\mathrm{peak}$ as function of the total binary mass $M_\mathrm{tot}$ for different EoSs and equal-mass mergers. Different EoSs are distinguished by different solid lines. The highest frequency $f^\mathrm{thres}_\mathrm{peak}$ for a given EoS is highlighted by a cross. The dashed line approximates the dependence of $f^\mathrm{thres}_\mathrm{peak}$ on the maximum total binary mass $M_\mathrm{stab}$ which still produces an (at least transiently) stable NS merger remnant. Right panel: Dominant GW frequency $f^\mathrm{thres}_\mathrm{peak}$ of the most massive NS merger remnant as a function of the radius $R_\mathrm{max}$ of the maximum-mass configuration of cold, nonrotating NSs for different EoSs (crosses). The diagonal solid line is a least-square fit to $f^\mathrm{thres}_\mathrm{peak}(R_\mathrm{max})$. Circles denote the estimated values for $f^\mathrm{thres}_\mathrm{peak}$ determined entirely from GW information from low-mass NS binary mergers (see text). Figures from \cite{Bauswein2014a}.}
  \label{fig:fpeakmtot}
\end{figure*}

Figure~\ref{fig:fpeakmtot} (left panel) displays the peak frequency as function of the total binary mass for different EoSs considering only equal-mass mergers. Each solid line corresponds to one particular EoS model. For a given EoS the dominant postmerger oscillation frequency increases continuously with the mass of the binary, which is expected since the compactness of the remnant increases with mass. Notably, in this diagram all sequences of the different EoSs terminate approximately on one particular curve (big crosses). This curve (dashed broken line) indicates the proximity to the threshold for prompt BH formation. Beyond this threshold no significant postmerger gravitational radiation is produced because the remnant directly collapses to a BH. Consequently, no $f_\mathrm{peak}$ can be measured. At the binary masses marked by big crosses the merger remnants are transiently stable and emit GWs for at least a short period. These binary masses are denoted as $M_\mathrm{stab}$ as the highest total binary mass which leads to a transiently stable NS remnant\footnote{Note that we use a slightly different nomenclature in \cite{Bauswein2014a}. $M_\mathrm{stab}$  corresponds to $M_\mathrm{thres}$ in \cite{Bauswein2014a}.}. The corresponding dominant frequency at $M_\mathrm{stab}$ is $f_\mathrm{peak}^\mathrm{thres}$. The frequency $f_\mathrm{peak}^\mathrm{thres}$ is the highest possible peak frequency that can be produced by a NS remnant. 

Determining $f_\mathrm{peak}^\mathrm{thres}$ and $M_\mathrm{stab}$ can provide additional constraints on the NS EoS that are detailed below and in Chapter~\ref{collapse}. Generally, an object on the brink to prompt collapse probes the very high density regime of the EoS. However, depending on the true value of $M_\mathrm{stab}$ a direct detection of $f_\mathrm{peak}^\mathrm{thres}$ may or may not be likely. Observations of binary NSs suggest that most mergers have a total binary mass of about 2.7~$M_\odot$ (see \cite{Lattimer2012,Oezel2016} for a compilation). If this result holds for the population of merging NSs, a binary merger with a total mass close to $M_\mathrm{stab}$ may be statistically less probable. In~\cite{Bauswein2014a} we proposed to employ two or more detections of $f_\mathrm{peak}$ of mergers with slightly different total binary masses in the most likely range of $M_\mathrm{tot}$ around 2.7~$M_\odot$. These measurements allow to estimate $f_\mathrm{peak}^\mathrm{thres}$ and $M_\mathrm{stab}$. This idea is based on the observation that $f_\mathrm{peak}(M_\mathrm{tot})$ depends in a continuous manner on $M_\mathrm{tot}$. Hence, two data points on the $f_\mathrm{peak}(M_\mathrm{tot})$ curve allow an extrapolation to higher $M_\mathrm{tot}$. Using a properly devised extrapolation procedure, one can determine $f_\mathrm{peak}^\mathrm{thres}$ and $M_\mathrm{stab}$, which are given by the intersection of the extrapolated $f_\mathrm{peak}(M_\mathrm{tot})$ curve and the dashed curve in Fig.~\ref{fig:fpeakmtot} (left panel). For details of this extrapolation method see \cite{Bauswein2014a}.

  \begin{figure*}
\begin{center}
\resizebox{0.9\textwidth}{!}{
\includegraphics{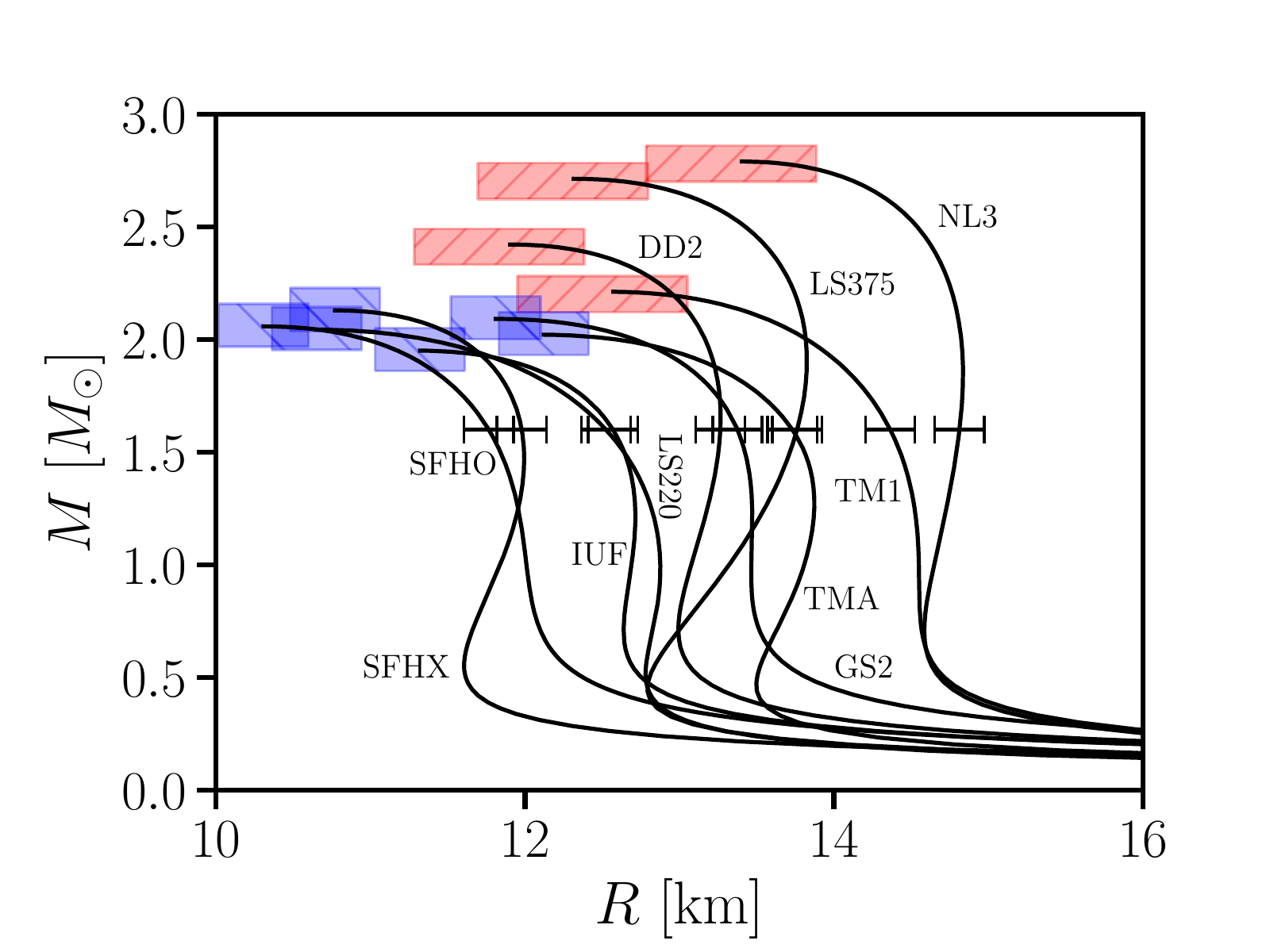}
}
\end{center}
  \caption{Mass-radius relations for different EoSs with the radius $R$ and the gravitational mass $M$. Boxes illustrate the maximum deviation of estimated properties of the maximum-mass configuration, which can be inferred from GW detections of low-mass binary NS mergers through an extrapolation procedure described in the text. The error bars indicate the accuracy of a radius measurement of a 1.6~$M_\odot$ NS.  Figure follows data in \cite{Bauswein2014a}, which can be consulted for further information.}
  \label{fig:rmboxes}
\end{figure*}

Estimating $f_\mathrm{peak}^\mathrm{thres}$ is important because the highest possible peak frequency, i.e.\ $f_\mathrm{peak}^\mathrm{thres}$, exhibits a tight correlation with the radius $R_\mathrm{max}$ of the maximum-mass configuration of nonrotating NSs. This relation is visualized in Fig.~\ref{fig:fpeakmtot} (right panel, Fig.~8 in \cite{Bauswein2014a}). Moreover, $f_\mathrm{peak}^\mathrm{thres}$ scales also with the maximum central rest-mass density of nonrotating NSs or the maximum central energy density of NSs (see Figs.~9 and~10 in \cite{Bauswein2014a}). An inversion of these relations determines these properties of nonrotating NSs similar to the radius measurements described above. For instance, the maximum densities in NSs can be estimated with a precision of about 10 per cent if $f_\mathrm{peak}^\mathrm{thres}$ is determined. An estimate of $M_\mathrm{stab}$ is particularly interesting because in combination with a radius measurement (see above) it provides a proxy for the maximum mass $M_\mathrm{max}$ of nonrotating NSs. Further details\footnote{In Sec.~\ref{collapse} we discuss the threshold binary mass $M_\mathrm{thres}$ for prompt collapse, which by definition is close to but slightly above $M_\mathrm{stab}$.} are provided in Sec.~\ref{collapse}. This may allow to determine $M_\mathrm{max}$ with an accuracy of $\pm 0.1~M_\odot$. The resulting EoS constraints are visualized in Fig.~\ref{fig:rmboxes} illustrating the precision to which stellar parameters of NSs may be determined.

In summary, the procedures developed in \cite{Bauswein2012,Bauswein2012a,Bauswein2014a,Clark2014,Bauswein2016,Clark2016,Chatziioannou2017} describe a way to determine stellar properties of NSs with high and moderate masses. As such, they provide a way to assess the very high density regime of the EoS, which cannot be probed by observing NS with lower masses. Since NSs with very high masses may be less frequent (see e.g.\ \cite{Oezel2016} for a compilation of measured masses), postmerger methods offer a unique way to understand properties of matter at the highest densities. We consider this one of the main advantages of detecting GWs from the postmerger phase of BNS mergers.

\section{Origin and interpretation of peaks in postmerger gravitational-wave spectra}
\label{secondary}

The GW spectrum of the postmerger phase of a NS merger exhibits many distinct peaks, e.g.~\cite{Stergioulas2011,Hotokezaka2013a,Takami2014,Kastaun2015,Bauswein2015,Palenzuela2015,Takami2015,Bauswein2016,DePietri2016,Foucart2016,Rezzolla2016,Clark2016,Dietrich2017a,2017CQGra..34c4001F,Maione2017}. Understanding the physical mechanisms generating these different features is essential for the detection and interpretation of postmerger GW signals. GW searches are more sensitive if additional information about the signal to be detected is available, for instance the general signal morphology. Therefore, it is important to comprehend the origin and the dependencies of the different structures of the postmerger GW spectrum. Moreover, the detection of several features of the postmerger phase bears the potential to reveal further detail of the incompletely known EoS of high-density matter beyond the constraints that can be obtained from a measurement of the dominant peak that we discussed extensively in Sec.~\ref{dominant}.  

Here, we discuss the nature of the dominant peak $f_{\rm peak}$ and explain the origin of the two most pronounced secondary peaks at lower frequencies, which we call $f_{\rm 2-0}$ and $f_{\rm spiral}$. Observationally, only the secondary peaks at frequencies smaller than $f_\mathrm{peak}$ are relevant, because the sensitivity of ground-based GW detectors decreases significantly at higher frequencies. Our findings are funneled into a unified picture of the postmerger dynamics and GW emission by considering a large number of merger simulations with different EoSs and total binary masses. We focus on equal-mass mergers and remark that small asymmetries lead to very similar results. The detailed impact on the secondary features of a strong asymmetry in the two masses still has to be worked out.

\begin{figure*}
\begin{center}
\resizebox{1.0\textwidth}{!}{
\includegraphics{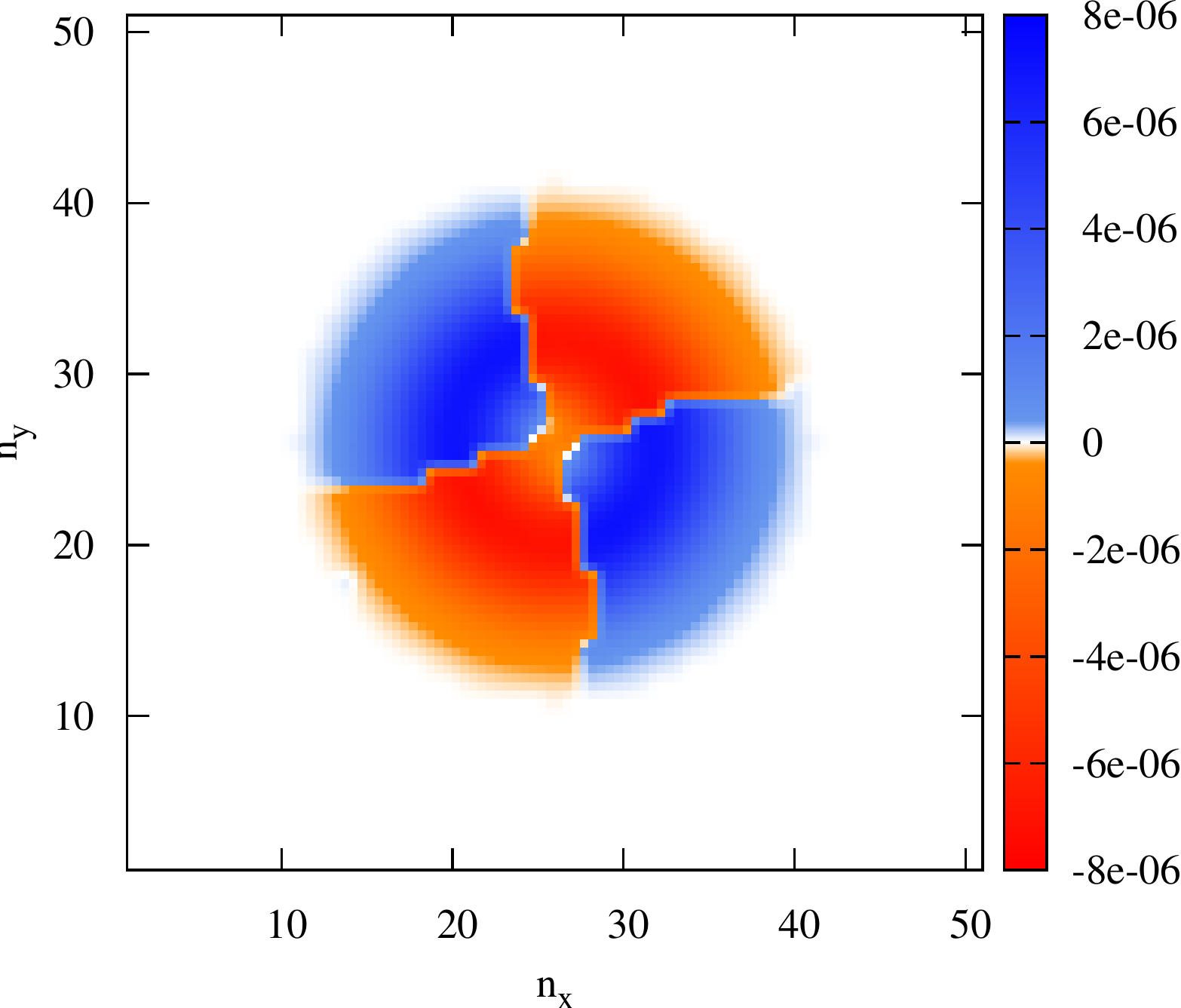}
  \hspace{4mm}
\includegraphics{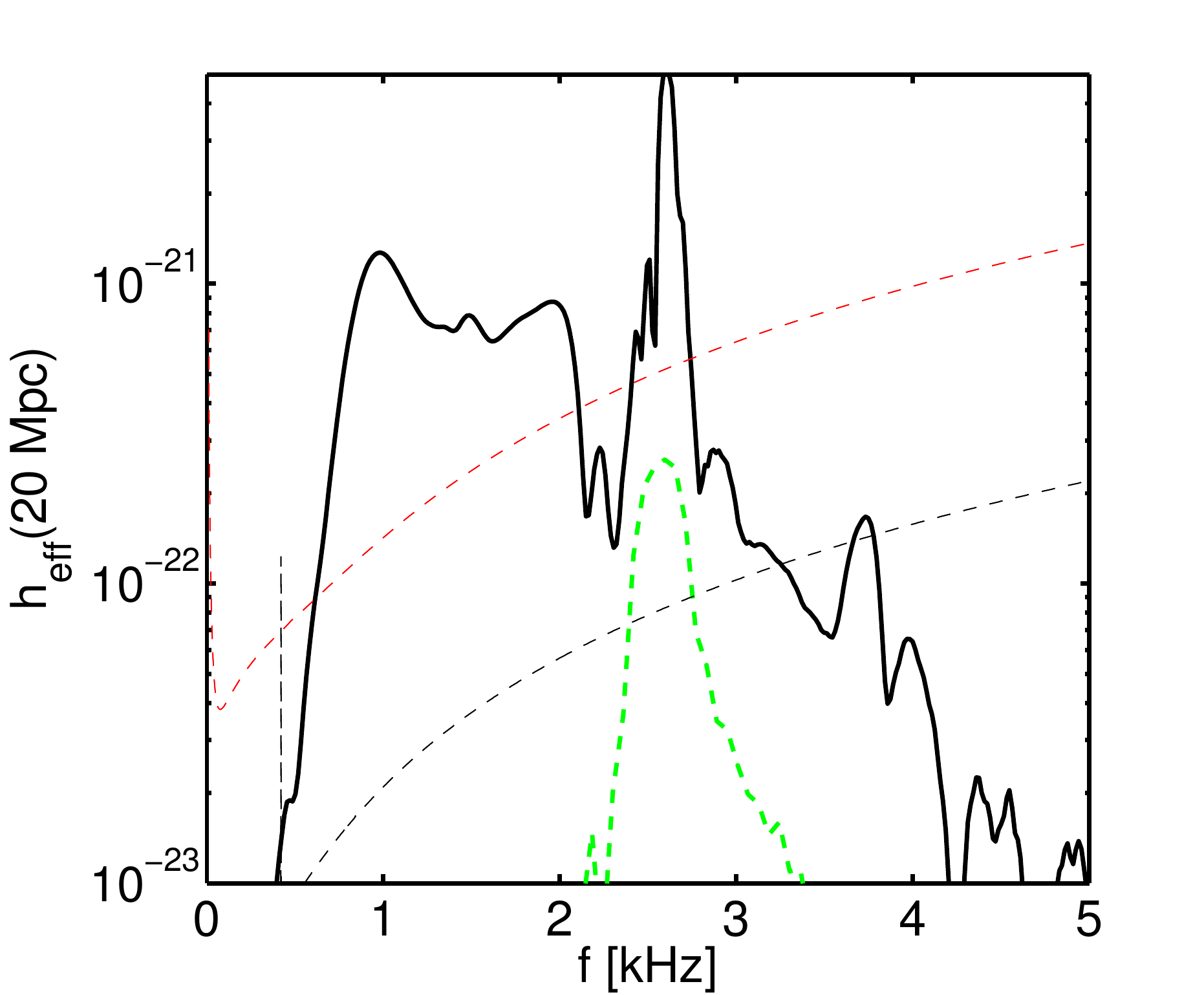}
}
\end{center}

  \caption{Left panel: Illustration of the eigenfunction of the pressure oscillation with a frequency $f=f_\mathrm{peak}$ in the equatorial plane for a 1.35-1.35~$M_\odot$ merger with the Shen EoS \cite{Shen1998}. Figure taken from \cite{Stergioulas2011}. Right panel: GW spectrum of a 1.35-1.35~$M_\odot$ merger with the DD2 EoS \cite{Hempel2010,Typel2010} (black line).  $h_\mathrm{eff}=\tilde{h}\cdot f$ with $\tilde{h}$ being the Fourier transform of the dimensionless strain $h(t)$ and $f$ is frequency.  The green dashed curve shows the GW spectrum of a simulation of a late-time merger remnant of the same model which was perturbed with a velocity field suitable to excite the fundamental quadrupolar fluid oscillation mode. Figures from \cite{Bauswein2016}.}
  \label{fig:fmode}
\end{figure*}

A powerful method for analyzing oscillation modes of rotating stars, based on a Fourier extraction of their eigenfunctions from simulation data, was presented in \cite{Stergioulas2004}. In \cite{Stergioulas2011} we applied this method for the first time to NS merger remnants. Figure~\ref{fig:fmode} elucidates the nature of the dominant feature in the postmerger GW spectrum by visualizing the eigenfunction of the mode with frequency $f_\mathrm{peak}$ (cf. Fig.~\ref{fig:spectrum}). The eigenfunction is extracted as follows: First, a Fourier analysis of the  evolution of pressure on a grid of fixed points covering the equatorial plane is performed (see Fig.~3 in \cite{Stergioulas2011}; alternatively the density evolution may be employed). Examining the Fourier spectra at the dominant frequency $f_\mathrm{peak}$ reveals that it is a discrete frequency throughout the star. Then, the Fourier amplitude at all points in the equatorial plane is extracted at the discrete frequency $f_\mathrm{peak}$. The resulting two-dimensional distribution of the amplitude represents the eigenfunction of this oscillation mode (the overall scaling is irrelevant, since the eigenfunctions are strictly defined as linear perturbations).  An example is shown color-coded in Fig.~\ref{fig:fmode} (left panel), which shows a clear quadrupolar structure (with an azimuthal mode number $m=2$) with no nodal lines in the radial direction. This analysis provides evidence that {\it the main peak in the GW spectrum is generated by the fundamental quadrupolar fluid mode}. By extracting the amplitude of the pressure oscillations at other frequencies, several other modes can be identified and associated with certain peaks in the GW spectrum or in the spectrum of the pressure evolution. This includes higher-order modes and quasi-radial oscillations (see Figs.~1 to~8 in \cite{Stergioulas2011}).

The above result is corroborated by additional hydrodynamical simulations of the late-time remnant, which settled into a quasi-stationary, nearly axisymmetric state. If an appropriate velocity perturbation is artificially added to the simulation at late times, one can specifically re-excite the fundamental quadrupolar fluid mode. Doing so, the perturbed remnant produces GW emission that strongly peaks at $f_\mathrm{peak}$ (see right panel of Fig.~\ref{fig:fmode}). This means that the frequency of the fundamental quadrupolar fluid mode of the remnant coincides with $f_\mathrm{peak}$. This provides further evidence that the main peak in the postmerger GW spectrum is generated by the fundamental quadrupolar fluid mode.

Considering the dynamics of the merger process, it is clear that the fundamental quasi-radial mode of the remnant is likely to be excited at a frequency which we denote as  $f_0$. Being nearly spherically symmetric, the quasi-radial mode produces only weak GW emission (normally at a frequency where the spectrum is still dominated by the inspiral phase). However, a non-linear coupling between the quasi-radial oscillation and the quadrupolar mode does emit strong GWs and explains some of the secondary peaks. At the lowest nonlinear interaction level, the coupling of the two modes  results in the appearance of {\it quasi-linear combination frequencies} $f_{2\pm0}=f_\mathrm{peak}\pm f_0$. Notice that the existence of such quasi-linear combination frequencies is a natural consequence of the nonlinear evolution of two different, simultaneous oscillations of the same star\footnote{In signal processing, the quasi-linear combination frequencies are an example of {\it second-order intermodulation products}, e.g. \cite{Signal2019}. In music theory, these were first described by Sorge in 1745 and by Tartini in 1754 and are known as {\it Tartini tones}, generated by nonlinearities, see \cite{Helmholtz}. } see e.g. \cite{2007PhRvD..75h4038P}.

To associate certain features in the GW spectrum with this mechanism, one needs to identify the quasi-radial mode $f_0$ from the hydrodynamical evolution, using the same Fourier technique as described above. It is also instructive to consider the time evolution of the central lapse function or the central density, which typically oscillate strongly and reflect the quasi-radial oscillation of the remnant (see e.g.\ Fig.~4 in \cite{Bauswein2015} or Fig.~10 in \cite{Bauswein2016}). Adding an artificial radial perturbation to the late-time remnant re-excites the quasi-radial mode and allows an unambiguous identification of the quasi-radial frequency $f_0$  in cases when the Fourier transform of the central lapse function exhibits two frequency peaks (see Fig. 10 in \cite{Bauswein2016}).

  \begin{figure*}
\begin{center}
\resizebox{1\textwidth}{!}{
\includegraphics{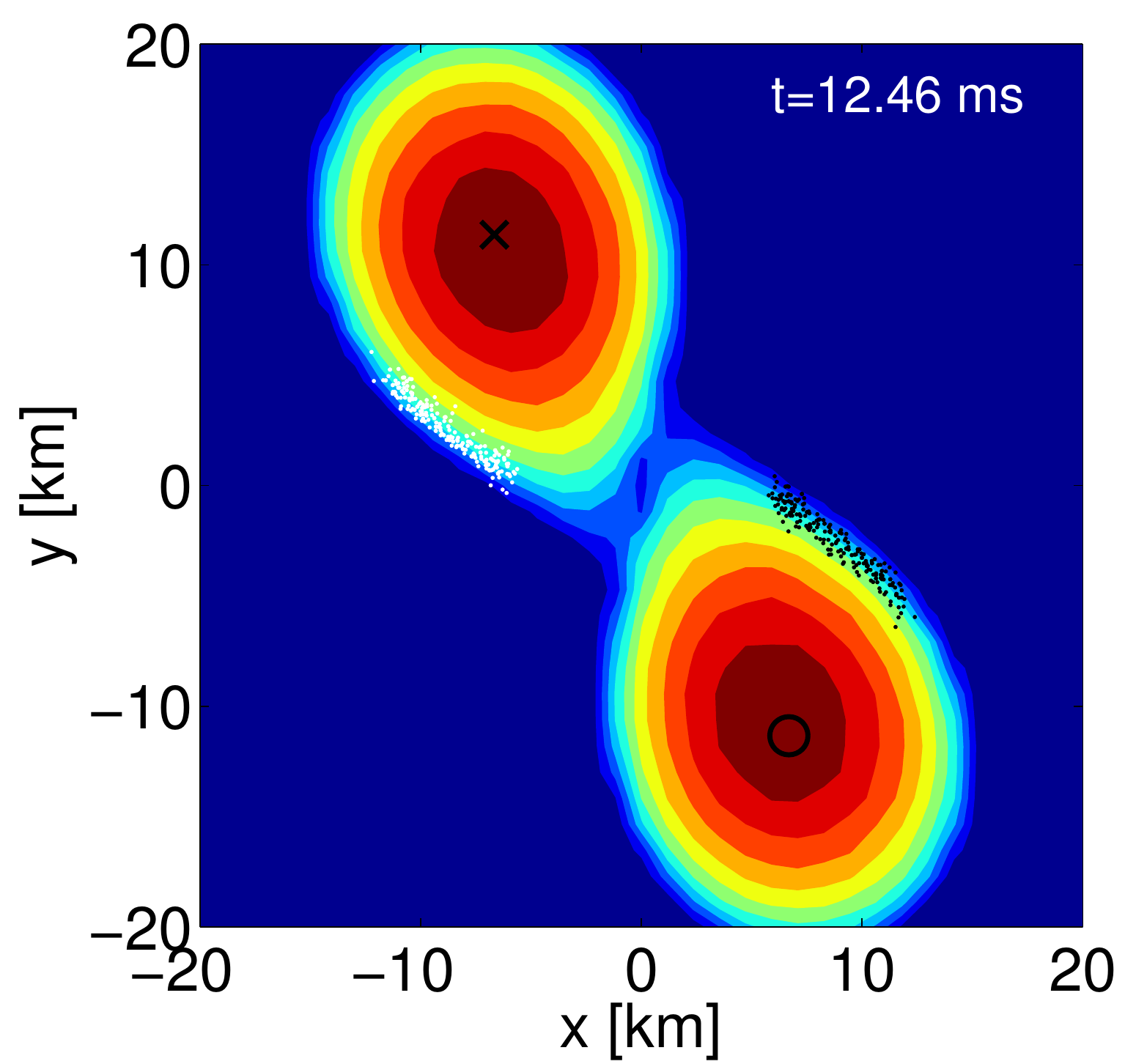}
\includegraphics{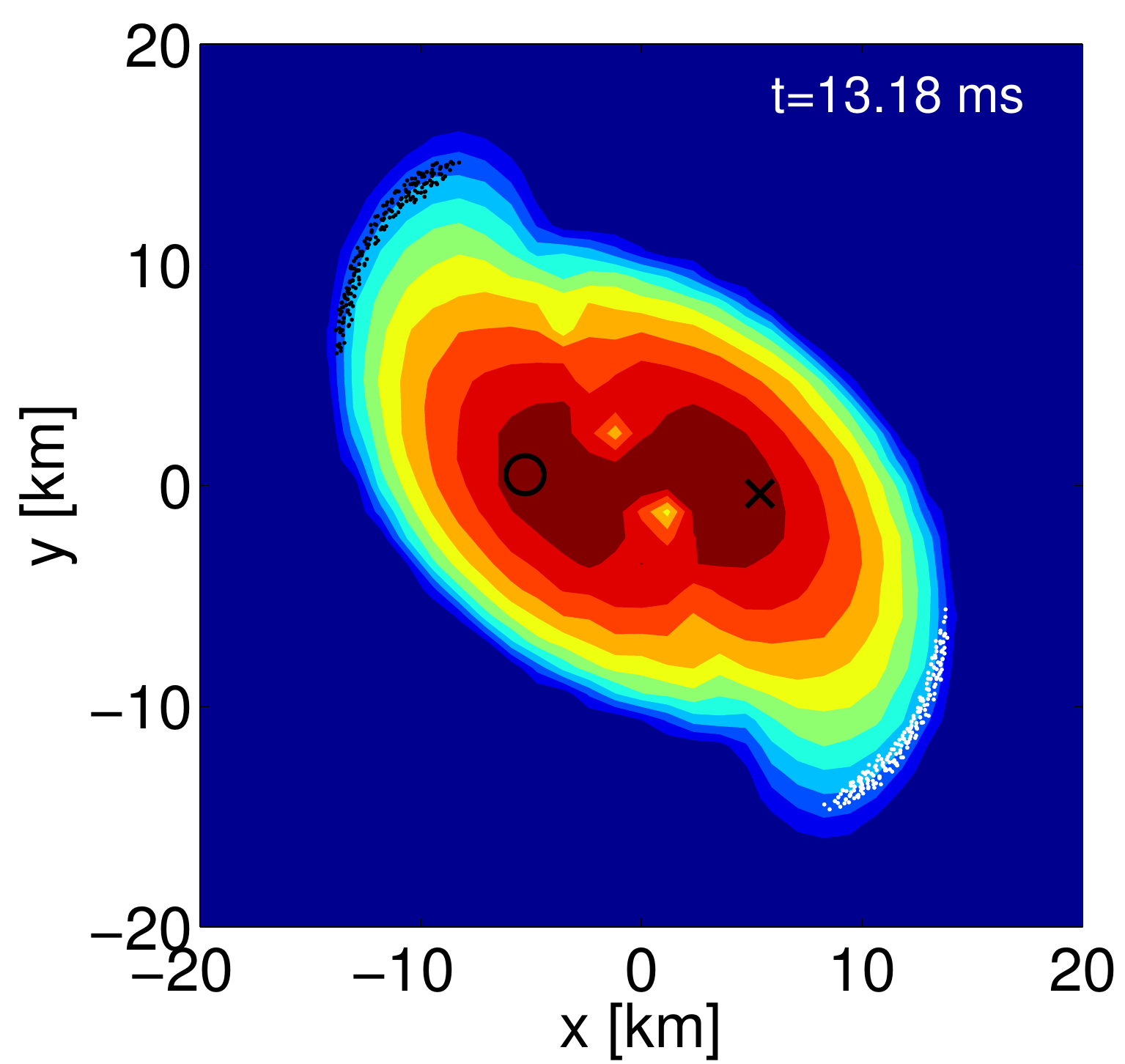}
\includegraphics{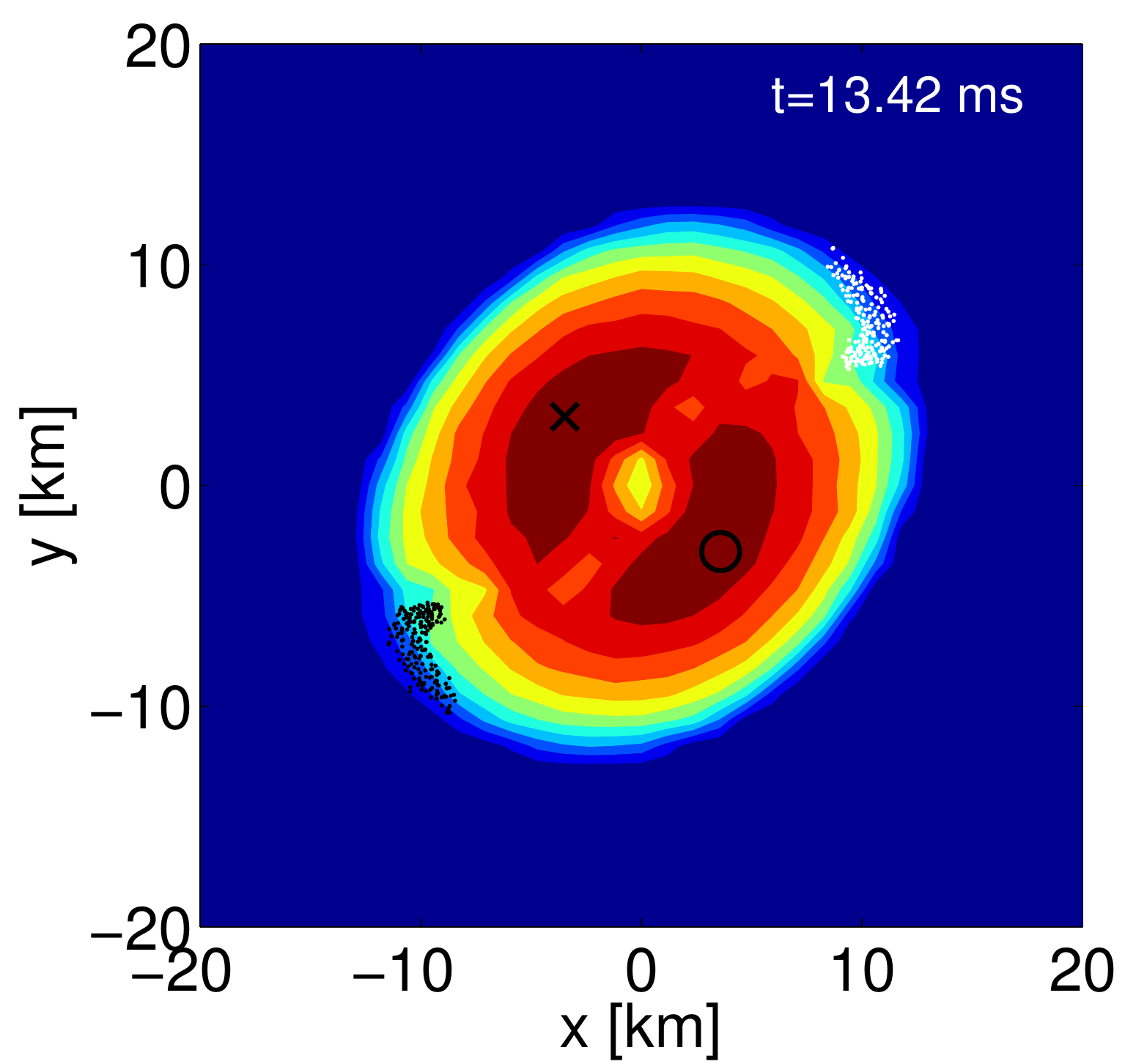}
\includegraphics{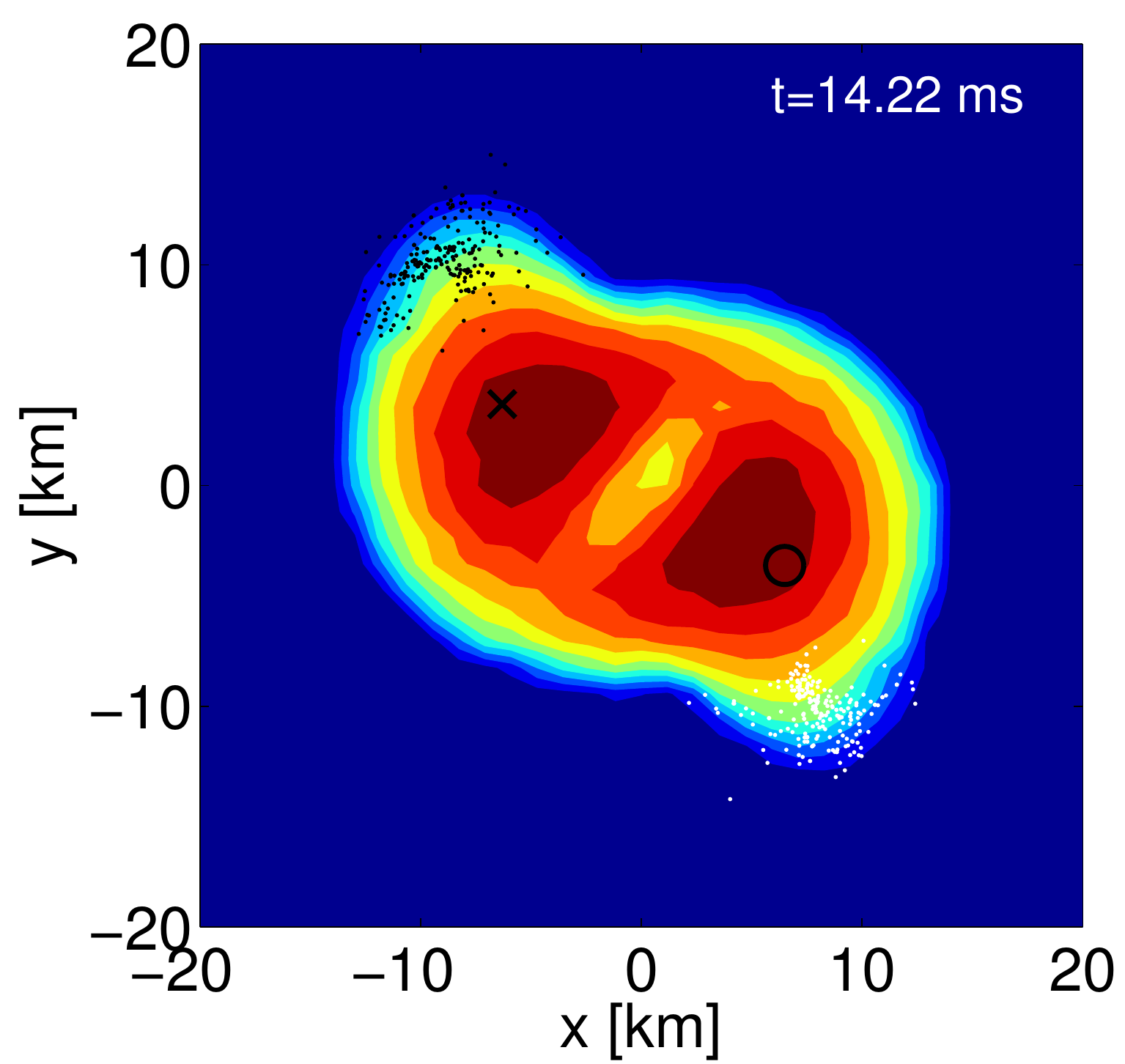}
}
\end{center}
  \caption{Rest-mass density evolution in the equatorial plane for the 1.35-1.35~$M_\odot$ merger with the DD2 EOS \cite{Hempel2010,Typel2010} (rotation counterclockwise). Black and white dots trace the positions of selected fluid elements of the antipodal bulges, which within approximately one millisecond complete one orbit (compare times of the right panels). The orbital motion of this pattern of spiral deformation produces the $f_\mathrm{spiral}$ peak in the GW spectrum at 2~kHz (see Fig.~\ref{fig:spectrum}). The cross and the circle mark the double cores, which rotate significantly faster than the antipodal bulges represented by the dots (compare times of the different panels). Figures from \cite{Bauswein2015}.}
  \label{fig:snap}
\end{figure*}

Once the $f_0$ mode and $f_{\rm peak}$ are determined, the corresponding secondary peaks at $f_{2\pm0}=f_\mathrm{peak}\pm f_0$ can be identified in the GW spectrum. For example, in the GW spectrum shown in Fig.~\ref{fig:spectrum} the $f_{2-0}$ feature is clearly visible at 1.5~kHz. The corresponding side peak at higher frequencies can be found at about 3.7~kHz. Remarkably, in some models the peaks at $f_\mathrm{peak}\pm f_0$ are strongly suppressed or even absent (see blue curve in Fig.~\ref{fig:class} (left panel)). In these systems the quasi-radial mode is not strongly excited, which then results in a suppression of the peaks at $f_{2\pm0}$ (see also blue curve in Fig.~4 of \cite{Bauswein2015}). This occurs predominantly for binary mergers with relatively low total binary masses and relatively stiff EoSs. On the other hand mergers with relatively high total binary masses and soft EoSs typically result in a strong excitation of the quasi-radial oscillation and consequently show a pronounced secondary peak at $f_\mathrm{peak}- f_0$ (see red curve in Fig.~\ref{fig:class} (left panel) and Fig.~4 of \cite{Bauswein2015}).

Having identified those secondary peaks that can be interpreted as the quasi-linear combination frequencies $f_{2\pm0}$, it becomes apparent that there is at least one more pronounced secondary peak at frequencies below $f_\mathrm{peak}$. This feature lies in between $f_{2-0}$ and $f_\mathrm{peak}$, see Fig.~\ref{fig:spectrum} and left panel of Fig.~\ref{fig:class}. In \cite{Bauswein2015} we provide evidence that this secondary peak is generated by the orbital motion of two bulges that form right after merging at the surface of the merger remnant (see Fig.~\ref{fig:snap}). During the merging the stars are strongly tidally deformed. Matter of this tidal deformation at the outer edges of the stars cannot follow the faster rotation of the cores of the original stars that constitute the inner part of the remnant. The material at the outer edges of the tidally stretched stars forms antipodal bulges, which orbit around the central remnant with a smaller orbital frequency (see dots in Fig.~\ref{fig:snap}). The structure dissolves within a few milliseconds, i.e.\ after about two revolutions. Because of the spiral-like pattern of the associated deformation in the case of equal-mass mergers (see upper right panel in Fig.~\ref{fig:snap}), we dubbed this feature in the GW spectrum as $f_\mathrm{spiral}$.

The following arguments support this picture. Extracting the orbital period of the bulges from the simulation data, the corresponding orbital frequency coincides with $f_\mathrm{spiral}/2$. For instance, in Fig~\ref{fig:snap} one orbit of the bulges is completed after about 1~ms (compare upper right and lower right panel), which would thus produce a peak at 2~kHz as seen in the GW spectrum (Fig.~\ref{fig:spectrum}). No other dynamical feature with this frequency can be identified in the hydrodynamical data. In particular, the highly deformed core of the remnant has a faster pattern speed than this. With an appropriate windowing of the GW signal, one can show that the $f_\mathrm{spiral}$ feature in the GW spectrum is produced within the first milliseconds after merging and thus coincides with the presence of the tidal bulges in the hydrodynamical data. From the hydrodynamical simulations the mass of the bulges can be estimated to amount to about 0.1~$M_\odot$ each. Within a simple toy model of two rotating point particles of 0.1~$M_\odot$ with an orbital frequency of $f_\mathrm{spiral}/2$, a GW peak with the observed amplitude is produced within a few orbits as seen in the simulations. Furthermore, it is found that the $f_\mathrm{spiral}$ feature is more pronounced in mergers with relatively low total binary masses and relatively stiff EoSs. This behavior is understandable because lower $M_\mathrm{tot}$ and stiff EoSs imply less bound stars, which favors the formation of massive tidal bulges.

Interestingly, the frequency difference $f_\mathrm{peak}-f_\mathrm{spiral}$ matches a frequency which is found in the time evolution of the central lapse function, but which does not agree with the frequency of the fundamental quasi-radial mode (see above and Fig.~4 in \cite{Bauswein2015}). This low-frequency modulation is very pronounced for low-mass NS mergers with relatively stiff EoSs (blue curve in Fig.~4 in \cite{Bauswein2015}). This modulation is naturally explained by the orientation of the bulges relative to the pattern of the quadrupolar deformation of the core. The orientation clearly affects the compactness of the whole system (lower compactness for aligned configuration, higher compactness for orthogonal configuration) and thus leaves an imprint on the evolution of the lapse function. The same low-frequency modulation can be observed in the time evolution of the central density and the GW amplitude of the postmerger phase.

\section{Spectral classification scheme}\label{sec:class}

Based on the understanding of the physical origin of the secondary GW peaks $f_{2-0}$ and  $f_\mathrm{spiral}$, an inspection of GW spectra of a large set of representative simulations leads to the following unified picture of the postmerger dynamics and GW emission, where one can distinguish three types of spectra and corresponding postmerger dynamics (notice that the dominant $f_\mathrm{peak}$ is present in all cases).

  \begin{figure*}
\begin{center}
\resizebox{1.0\textwidth}{!}{
\includegraphics{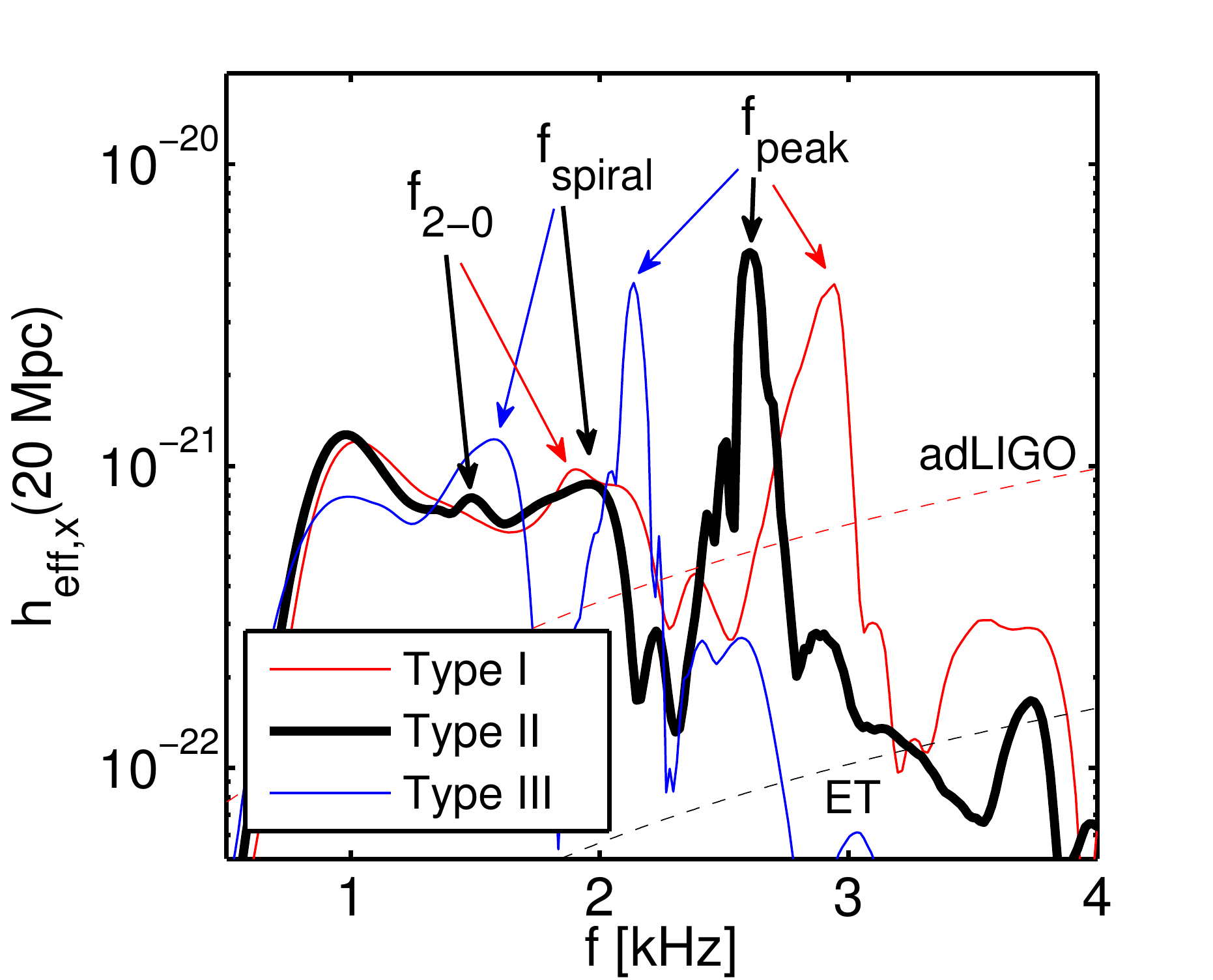}
\includegraphics{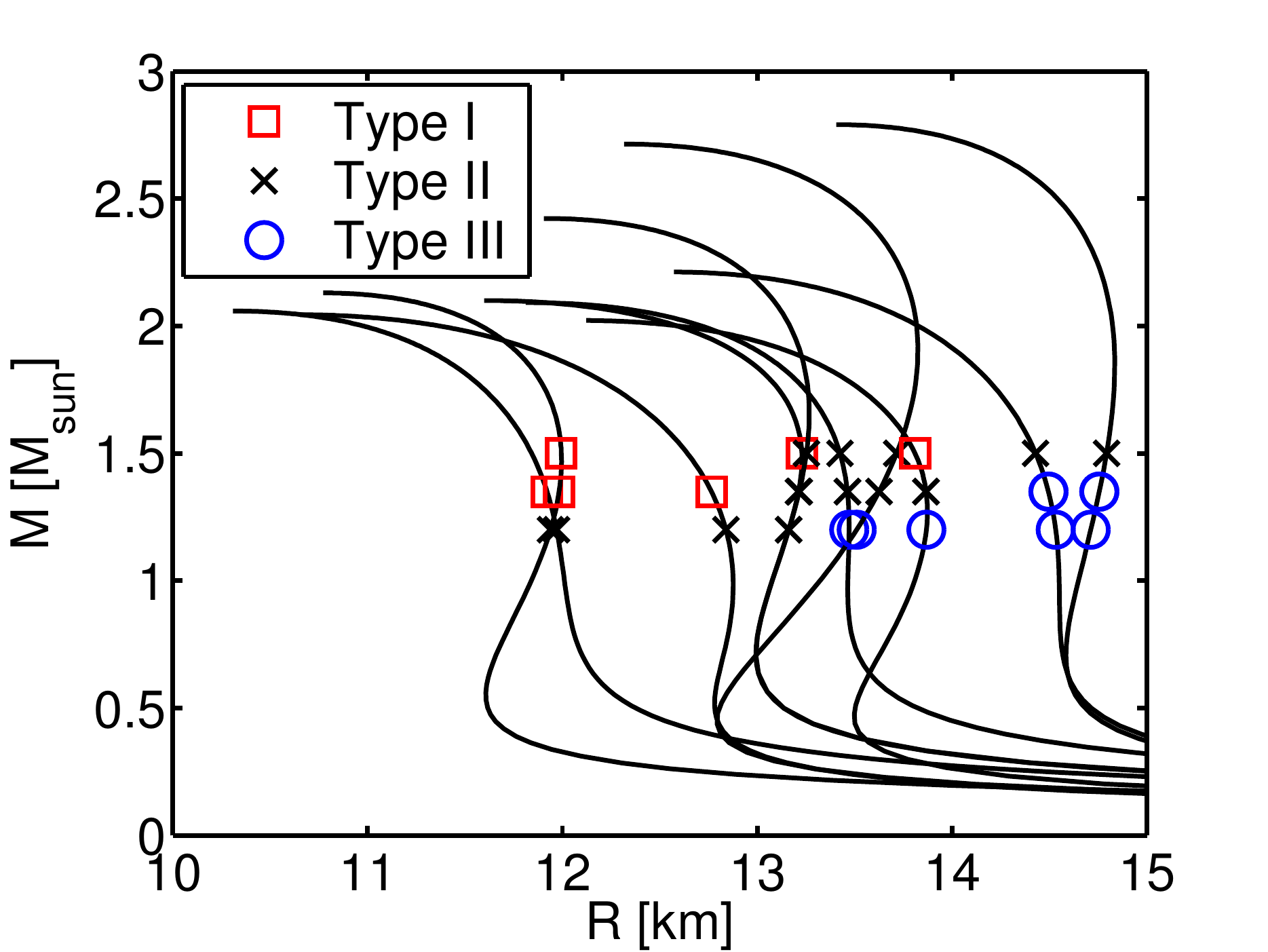}
}
\end{center}

  \caption{Different types of postmerger dynamics and GW emission of different merger models. Left panel shows GW spectra of the different types 
  ($h_\mathrm{eff}=\tilde{h}\cdot f$, with $\tilde{h}$ being the Fourier transform of the dimensionless strain $h(t)$ and $f$ is frequency).  Right panel surveys a large sample of simulations. The outcome of a given calculation with a total binary mass $M_\mathrm{tot}$ is shown as a symbol at $M_\mathrm{tot}/2$ plotted on the mass-radius relation of the EOS employed in the respective simulation. Red squares indicate type I, black crosses stand for type II, and blue circles mark type III. See text for definitions of different types of postmerger dynamics and GW emission. Figures  from \cite{Bauswein2015}.}
  \label{fig:class}
\end{figure*}

\begin{itemize}
\item {\bf Type I}: The $f_{2-0}$ peak is the strongest secondary feature, while the $f_\mathrm{spiral}$ peak is suppressed or hardly visible. The time evolution of the central lapse function and the maximum density show a very clear and strong oscillation with a single frequency $f_0$, which corresponds to the fundamental quasi-radial mode of the remnant. This behavior is found for mergers with relatively high total binary masses and soft EoSs. In these cases, the individual NSs before merger are more compact and more tightly bound, and the collision occurs with a high impact velocity (see Fig.~3 in \cite{Bauswein2013a}) since the inspiral lasts somewhat longer. Consequently, the quasi-radial mode is strongly excited, leading to a pronounced $f_{2-0}$ feature, whereas the formation of tidal bulges is suppressed because of the stronger binding, which explains the weakness of the $f_\mathrm{spiral}$ peak.

\item {\bf Type II}: In this spectral type, both secondary features $f_{2-0}$ and $f_\mathrm{spiral}$ are clearly present and have roughly comparable strength. Moreover, the time evolution of the central lapse function and of the maximum density exhibit two main  frequencies: the fundamental quasi-radial mode $f_0$ and a modulation with $f_\mathrm{peak}-f_\mathrm{spiral}$ as explained above (see e.g.\ Fig.~10 in \cite{Bauswein2016}). These features of the GW emission and postmerger dynamics are observed for followning combination of parameters: very low masses and soft EOS; average masses and EoS of average stiffness;  high masses and stiff EoS.

\item {\bf Type III}: Low-mass mergers in particular with stiff EoSs produce GW spectra where the $f_\mathrm{spiral}$ feature is the most prominent secondary peak, whereas a peak at $f_{2-0}$ is either strongly suppressed or even absent. The time evolution of the central lapse function and the maximum density are dominated by the $f_\mathrm{peak}-f_\mathrm{spiral}$ modulation. This frequency is also visible as a modulation in the amplitude of the postmerger GW signal. These observations are readily explained by the merging of less compact NSs. The merger process proceeds in a relatively smooth manner with a relatively small impact velocity in comparison to Type I mergers. Therefore, the quasi-radial oscillation of the remnant is not strongly excited and thus hardly visible in the time evolution of the central lapse function. This weaker excitation of the quasi-radial mode also explains the absence of the $f_{2-0}$ peak. The smaller NS compactness favors the formation of massive tidal bulges and thus a very pronounced $f_\mathrm{spiral}$ peak.
\end{itemize}

Figure~\ref{fig:class} (right panel) illustrates the classification of postmerger spectra from different simulations according to the classification scheme presented above. The type of a merger with a given EoS and $M_\mathrm{tot}$ is displayed on the mass-radius relation of the employed EoS (using the mass of the individual binary components, i.e.\ at $M_1=M_2=M_\mathrm{tot}/2$). The diagram clearly visualizes the existence of the different types depending on the NS compactness and the total binary mass, as described above. We stress that for a given EoS different spectral types may occur, depending on the total mass of the binary: Type I for mergers with binary masses close to the threshold mass for prompt BH formation, Type II for mergers with intermediate $M_\mathrm{tot}$, and Type III for mergers with relatively low total binary mass. The terms ``relatively high'' or ``relatively low'' total binary mass are meant with respect to the EoS-dependent threshold binary mass for prompt collapse. Thus, the association of a given $M_\mathrm{tot}$ as being relatively high or low depends on the EoS. In any case, the classification of a given GW spectrum according to these criteria is only tentative as the transition between the different classes is continuous. 

In the range of total masses $2.4 M_\odot \leq M_{\rm tot} \leq 3.0 M_\odot$, the secondary peaks appear in distinct frequency ranges: $f_{\rm peak} - 1.3 {\rm kHz} \leq f_{2-0} \leq f_{\rm peak} - 0.9 {\rm kHz} $, while $f_{\rm peak} - 0.9 {\rm kHz} \leq f_{\rm spiral} \leq f_{\rm peak} - 0.5{\rm kHz} $. This 
property will be useful for identifying either $f_{2-0}$ or $f_{\rm spiral}$ (or both) in future GW observations.

We should note that for asymmetric mass ratios of $q\sim 0.7$ the above classification scheme has to be modified, to accommodate for the somewhat different postmerger dynamics with respect to the equal-mass case (for example, in such asymmetric cases the tidal deformation will not be a symmetric spiral and this will considerably weaken the $f_{\rm spiral}$ secondary peak).

\section{EoS dependence of secondary gravitational-wave peaks}\label{sec:seceos}

Similar to the dominant postmerger GW frequency, the frequencies of the secondary peaks in the GW spectrum depend in a particular way on the EoS. This is important because a measurement of these frequencies could provide additional information on the high-density EoS. However, one should keep in mind that these frequencies are more difficult to be detected, because the secondary peaks typically have smaller SNR,  in comparison to the main peak. Moreover, the secondary peaks can be relatively broad, which may lead to larger errors in there determination, as compared to the dominant peak at $f_\mathrm{peak}$.

\begin{figure*}
\begin{center}
\resizebox{1.0\textwidth}{!}{
\includegraphics{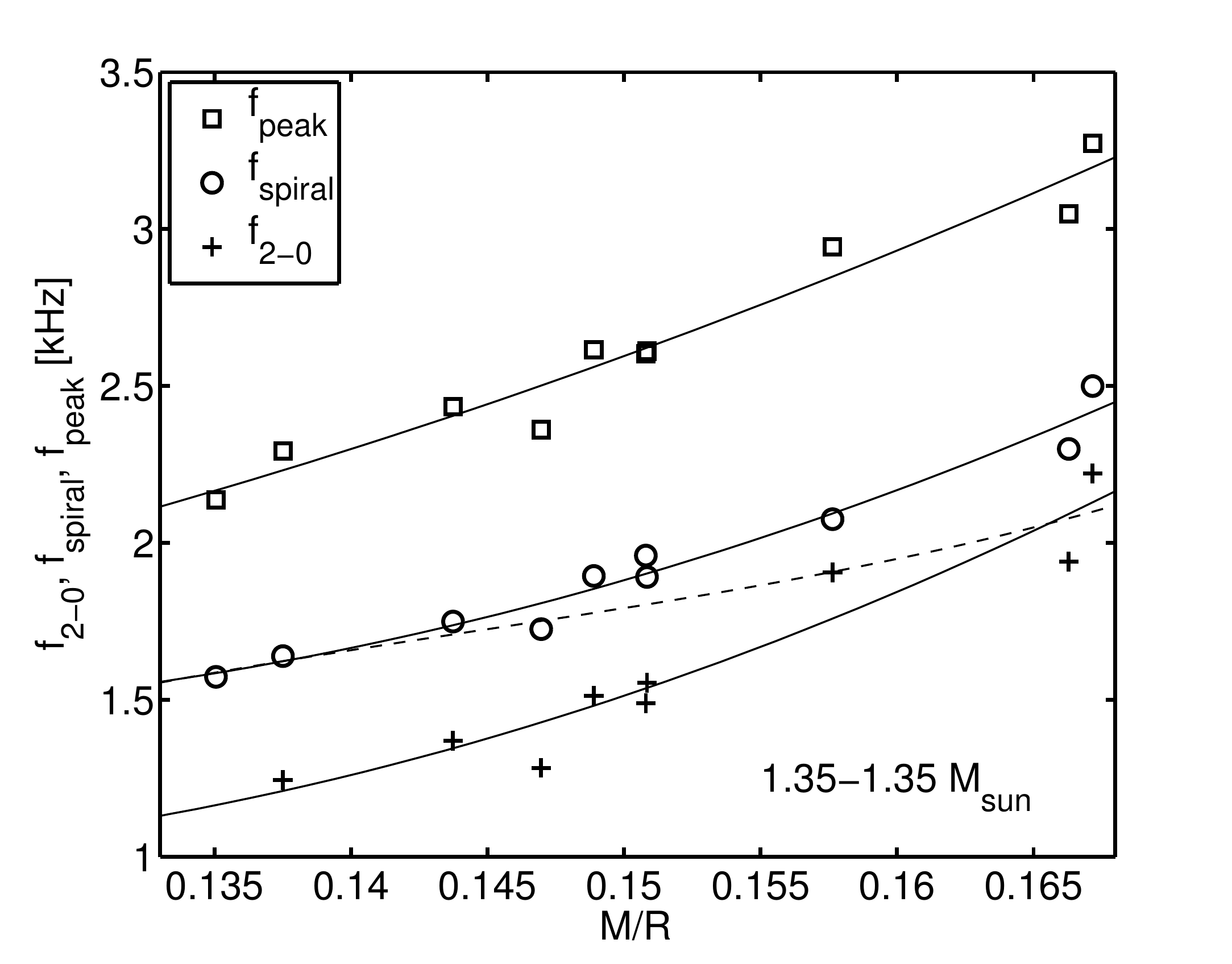}
\includegraphics{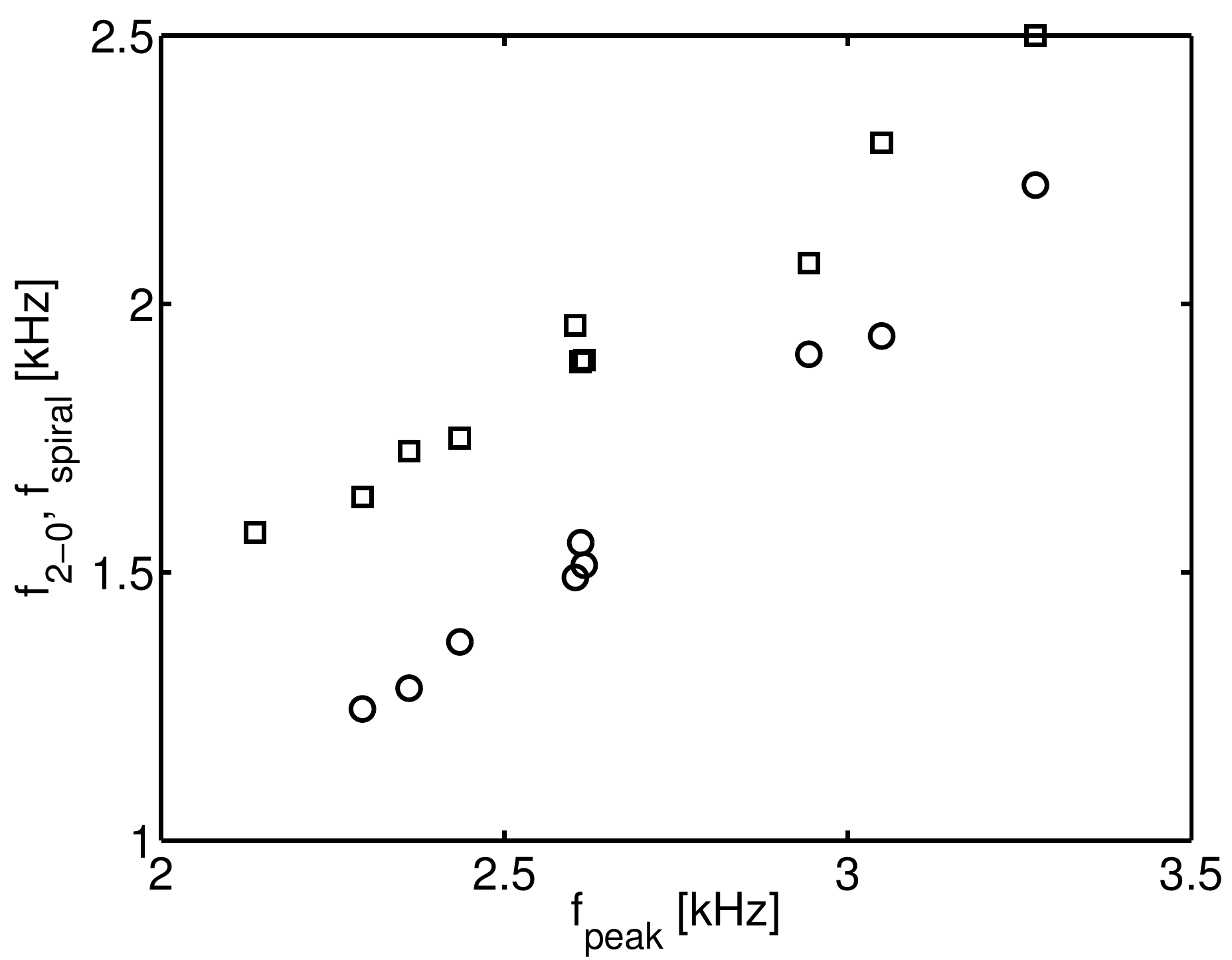}
}
\end{center}
  \caption{Left panel: $f_\mathrm{peak}$, $f_\mathrm{spiral}$ and $f_{2-0}$ for equal-mass mergers with ten different EoSs and $M_\mathrm{tot}=2.7~M_\odot$ versus the compactness $\frac{G\,M}{c^2\,R}$ of nonrotating NSs with a mass of $M=1.35~M_\odot$. Solid lines show empirical relations. The dashed line is taken from \cite{Takami2014}. Figure taken from \cite{Bauswein2015}. Right panel: Secondary frequencies $f_{2-0}$ (circles) and $f_\mathrm{spiral}$ (squares) as function of the dominant postmerger GW frequency $f_\mathrm{peak}$ for 1.35-1.35~$M_\odot$ mergers with different EoSs. Figure taken from \cite{Clark2016}.}
  \label{fig:secfreq}
\end{figure*}

Following a similar strategy as for the main peak, we explore relations between the frequencies of the subdominant peaks and stellar properties of nonrotating NSs. Since individual masses can be assumed to be measured sufficiently well from the insprial GW signal, we focus on sets of simulations for fixed total binary masses. Figure~\ref{fig:secfreq} (left panel) shows the three characteristic postmerger GW frequencies $f_\mathrm{peak}$, $f_\mathrm{spiral}$ and $f_{2-0}$ for 1.35-1.35~$M_\odot$ mergers for different EoSs as a function of the compactness ${\cal C}={G\,M}/{c^2 R}$ of nonrotating NS with 1.35~$M_\odot$. Note that for fixed masses the NS compactness is fully equivalent to NS radii, which are used in Sec.~\ref{dominant} to characterize $f_\mathrm{peak}$. There is a clear hierarchy of the GW frequencies as follows: $f_\mathrm{peak} > f_\mathrm{spiral} > f_{2-0}$. Figure~\ref{fig:secfreq} also reveals that all frequencies scale in a similar way with the compactness. Simulations with EoSs which lead to more compact NSs, yield higher frequencies. This may not be unexpected given the physical mechanisms forming these different features. Moreover, we found in \cite{Bauswein2015} that very similar scalings with the compactness exist also for other fixed binary masses, e.g.\ 1.2-1.2~$M_\odot$ and 1.5-1.5~$M_\odot$ mergers (see Figs.~7 and~8 in \cite{Bauswein2015}). 

Our study has also clarified whether a universal mass-independent relation for a single subdominant peak exists as claimed in \cite{Takami2014,Takami2015,Rezzolla2016}. In those publications, no distinction between the two different secondary peaks was made.  Their claimed relation is shown as dashed curve in Fig.~\ref{fig:secfreq} (left panel) and clearly does not reproduce the full spectrum of secondary peaks in the GW spectrum. In the light of our findings it seems likely that their claimed universal relation describes selectively either $f_{2-0}$ or $f_\mathrm{spiral}$ in different regions of compactness. For lower NS compactnesses $f_\mathrm{spiral}$ is dominant, whereas EoSs leading to more compact NSs yield a relatively pronounced $f_{2-0}$ peak fully in line with the unified picture described above (Sec.~\ref{sec:class}, Ref.~\cite{Bauswein2015}). Thus, \cite{Takami2014} may have picked selectively the strongest secondary peak, which however is clearly ambiguous in cases where both subdominant features have comparable strength (Type II in our classification scheme). In those cases, using a single universal relation for secondary peaks is certain to fail. Based on our simulations with a larger set of candidate EoSs we thus do not confirm the existence of a universal relation. Importantly, the claimed relation does not reproduce the full data for different total binary masses if $M_\mathrm{tot}$ varies within a representative range between 2.4 and 3.0~$M_\odot$ (see dashed curves in Figs.~7 and~8 in \cite{Bauswein2015}). Instead of following a single mass-independent relation (as claimed in \cite{Takami2014})  our set of simulations clearly shows that the secondary frequencies fulfill different scalings with the NS compactness for different fixed binary masses. We emphasize that there is no conflict between the data of the different groups for specific binary setups. The different findings are a result of the different sets of simulated binary setups (EoSs, $M_\mathrm{tot}$). Our findings are based on a larger set of candidate EoSs and assumed masses. The overview panels in~\cite{Rezzolla2016} showing GW spectra of different simulations are fully compatible with our picture: The overplotted frequencies of the secondary peaks as predicted by the fit formulas in~\cite{Bauswein2015} agree remarkably well with these independent calculations taking into account that the different empirical relations for $f_{2-0}$ and $f_\mathrm{spiral}$ exhibit a sizable intrinsic scatter (see Fig.~\ref{fig:secfreq}) and that~\cite{Rezzolla2016} employed a different set of EoSs where temperature effects have been treated in an approximate manner for most models.

We also rule out the interpretation of the secondary GW peaks being produced by a single instantaneous GW frequency, which strongly varies in time, as proposed in \cite{Kastaun2015,Takami2015}. Within this picture the different secondary peaks are claimed to be produced at the different extrema of the instantaneous GW frequency. We do not follow this interpretation for three reasons. First, spectrograms show no evidence for a very strong time variation of a single frequency but instead clearly reveal the presence of {\it several} distinct frequencies (see \cite{Clark2016}). Second, the idea is in conflict with the fact that a Fourier transform of only the early postmerger signal is dominated by the peak at $f_\mathrm{peak}$. This is not expected if power in the spectrum accumulates predominantly at the extrema of the instantaneous GW frequency. Third, this interpretation has not been corroborated by quantitative arguments as in our analysis described above. In fact, the hydrodynamical simulation data do not exhibit any dynamical feature that would be compatible with the claimed mechanism of a rapidly changing single (orbital) frequency of the remnant. Instead, a strongly varying instantaneous GW frequency is naturally explained as a synthesis of several slowly-varying distinct frequencies, each of which has a different physical origin, as explained above.

Finally, it is instructive to plot the frequencies of the secondary peaks as function of $f_\mathrm{peak}$ as in Fig.~\ref{fig:secfreq} (right panel). The frequencies $f_\mathrm{spiral}$ and $f_{2-0}$ scale tightly with the dominant frequency, which implies that the different frequencies encode very similar information about the EoS. On the other hand, these scalings explain why GW spectra show a certain universality (see discussion and Fig.~5 in \cite{Clark2016} and Fig.~2 in \cite{Bauswein2016a}). These findings finally provide an explanation why a principal component analysis can be successfully employed for GW data analysis \cite{Clark2016}. Moreover, the understanding of the different mechanisms shaping the postmerger GW signal can be funnelled into an analytic model that is able to reproduce simulation data very well (see Figs.~12 to~14 in \cite{Bauswein2016}).  
 
\section{Collapse behavior and EoS constraints} \label{collapse}

The dynamics and the outcome of a NS merger are mostly determined by the total binary mass and the assumed EoS. After describing the postmerger GW emission, we discuss now the collapse behavior of NS mergers, i.e.\ the distinction between a prompt BH collapse and the formation of metastable NS merger remnant. A NS merger remnant may or may not form a BH after a dynamical or secular postmerger evolution. Several physical mechanisms act to potentially induce a ``delayed collapse'' such as GW emission, mass loss, angular momentum redistribution, neutrino cooling and magnetic spin down. Here, we do not further distinguish between a delayed or no collapse, keeping in mind that the determination of the lifetime of a NS merger remnant and its long-term evolution is very challenging, because all mentioned physical effects have to be adequately modeled. Such efforts are in particular limited by available computational power. Also, it may be difficult to measure the exact lifetime in an observation.

Considering the occurrence of a prompt BH collapse is interesting, because it can be observationally discriminated from a NS remnant. If a NS postmerger remnant forms, it produces strong GW emission in the frequency range between 2 and 4~kHz. In contrast, the gravitational radiation of a promptly forming BH is significantly weaker and peaks at higher frequencies \cite{Shibata2006,Baiotti2008}. The GW emission of a NS remnant is the strongest during the first milliseconds after merging such that also remnants with short lifetimes produce significant emission and can be distinguished from a prompt collapse. In \cite{Clark2014} it has been shown that GW data analysis based on an unmodeled search can discriminate a prompt collapse from a delayed collapse for sufficiently loud events. Moreover, the collapse behavior has a crucial impact on the amount of dynamical ejecta. A prompt collapse event produces significantly less ejecta (see Fig.~7 in \cite{Bauswein2013a}) and therefore leads to a dimmer electromagnetic counterpart \cite{Hotokezaka2013,Bauswein2013a}. We recall that the GW inspiral signal reveals the  total binary mass with good accuracy, while the mass ratio and the distance to the source are measured with somewhat worse precision.

\section{Binary threshold mass for prompt black-hole formation}\label{sec:thres}

It is intuitively clear and confirmed by simulations that a direct BH collapse occurs for higher total binary masses while mergers with smaller total binary masses lead to NS remnants. Hence, there exists a threshold binary mass $M_\mathrm{thres}$ that separates the prompt gravitational collapse and the formation of a NS remnant. Since total binary masses are measurable from the GW inspiral signal, the threshold binary mass can be determined through the different observational signatures of the merger outcomes as sketched above. Moreover, it is clear that $M_\mathrm{thres}$ depends sensitively on the assumed EoS of NS matter, in the same way that the stability of nonrotating NS is sensitive to the EoS.  \cite{Shibata2005,Hotokezaka2011,Bauswein2013}. Understanding the EoS dependence of $M_\mathrm{thres}$ is the main purpose of the studies summarized in this section. A measurement of $M_\mathrm{thres}$ may in turn reveal unknown properties of the high-density EoS.

   \begin{figure*}
\begin{center}
\resizebox{1.0\textwidth}{!}{
\includegraphics{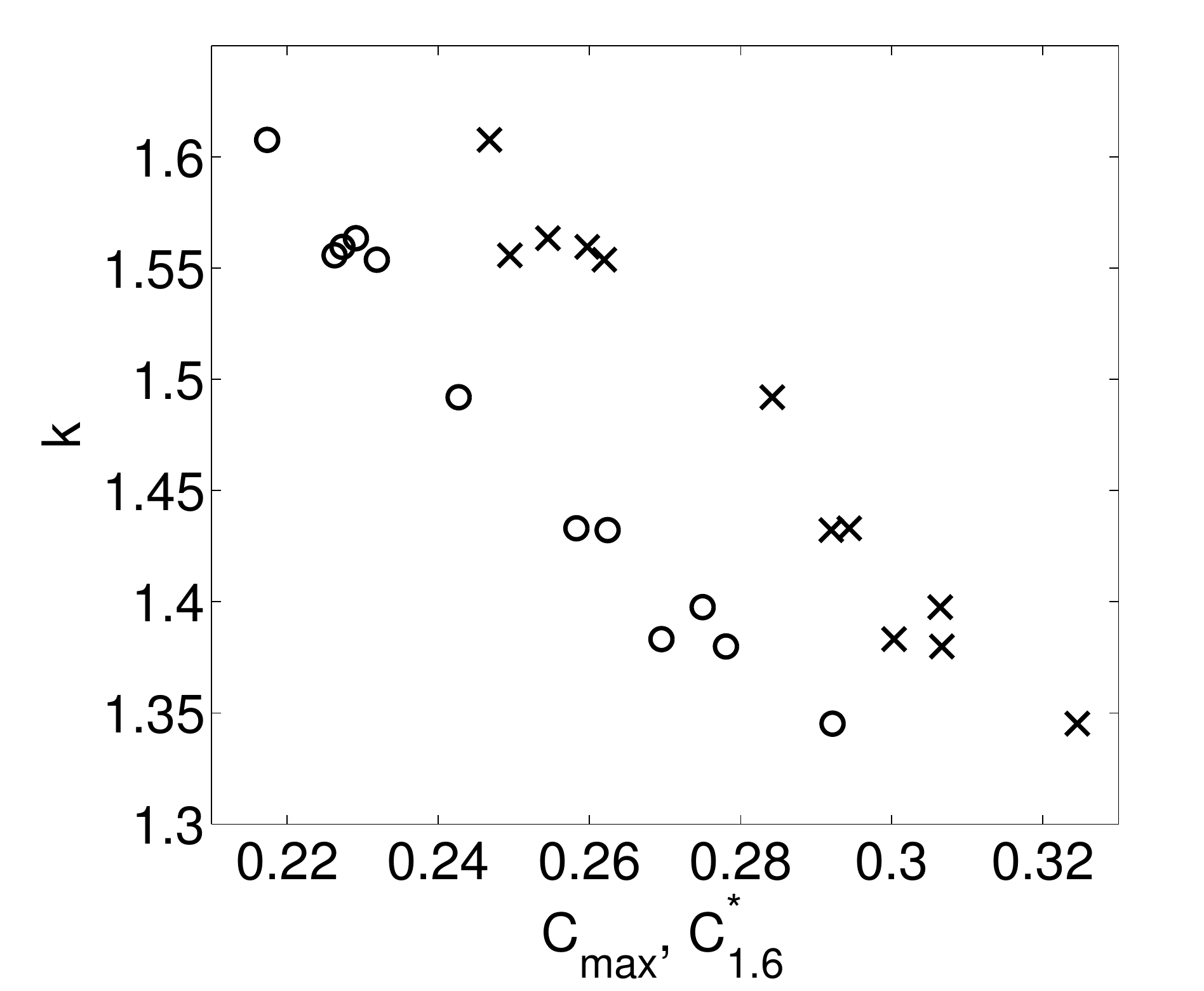}
\includegraphics{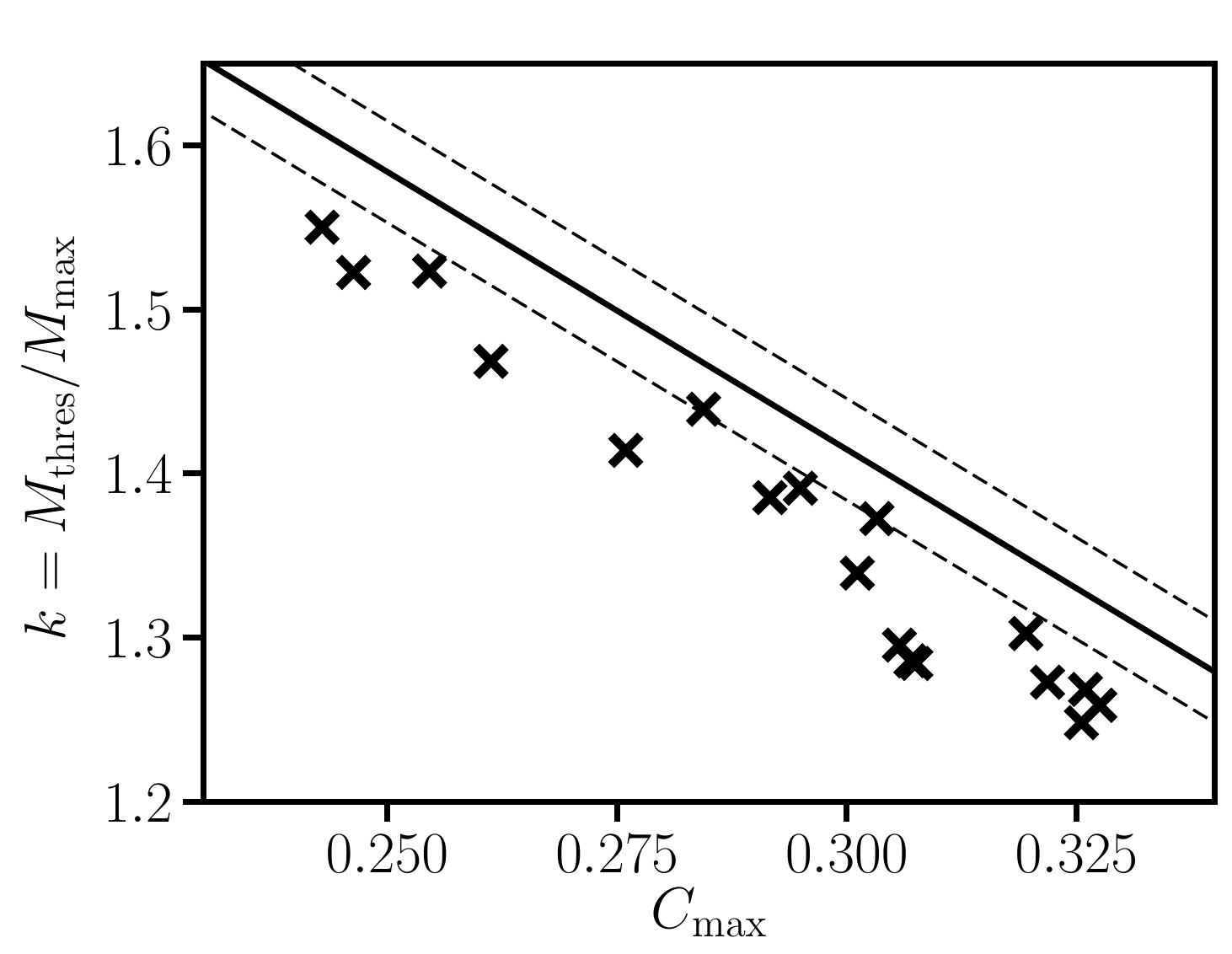}
}
\end{center}
  \caption{Left panel: Ratio $k$ between the binary threshold mass for prompt BH formation and the maximum mass $M_\mathrm{max}$ of nonrotating NSs as function of the compactness $C_\mathrm{max}=\frac{G\,M_\mathrm{max}}{c^2\,R_\mathrm{max}}$ of the maximum-mass configuration of nonrotating NSs (crosses). Circles display $k$ as function of $C^*_\mathrm{1.6}={G\,M_\mathrm{max}}/{c^2\,R_{1.6}}$. Figure from \cite{Bauswein2013}. Right panel: Ratio $k$ computed from a semi-analytic model (see text) as function of the compactness $C_\mathrm{max}$ (crosses). The solid line is a least-square fit to the data from dynamical merger simulations displayed by crosses in the left panel. The dashed curves indicate the deviations of the simulation data from the fit. Figure taken \cite{Bauswein2017a}.}
  \label{fig:kcmax}
\end{figure*}

The threshold binary mass is determined by a large set of simulations for different candidate EoSs. For a given EoS, simulations are performed for different total binary masses and the corresponding $M_\mathrm{thres}$ is found by checking the outcome of the different computations. The highest $M_\mathrm{tot}$ leading to a NS remnant and the lowest $M_\mathrm{tot}$ resulting in a prompt collapse determine $M_\mathrm{thres}$ with an accuracy that is given by the chosen sampling in the $M_\mathrm{tot}$ space. In \cite{Bauswein2013} we focus on equal-mass mergers and compute the threshold binary mass for a set of 12 representative EoSs with an accuracy of $\pm 0.05~M_\odot$ (see Tab.~1 therein). Additional simulations for selected EoSs show that asymmetric binary systems lead to the same threshold mass if the mass ratio is not too extreme.

Similarly to the approach in Sec.~\ref{dominant}, we investigate the EoS dependence by searching for empirical relations between the observable quantity, here $M_\mathrm{thres}$, and stellar properties of nonrotating NSs as characteristics of the given EoSs. It is convenient to introduce the ratio $k=M_\mathrm{thres}/M_\mathrm{max}$, which quantifies the fraction of the maximum mass of nonrotating NSs that can be supported against the gravitational collapse in a hot, rotating merger remnant by the stabilizing effects of rotation and thermal pressure. 

Figure~\ref{fig:kcmax} (left panel) shows a clear relation between $k$ and the maximum compactness of nonrotating NSs being defined by $C_\mathrm{max}=\frac{G\,M_\mathrm{max}}{c^2\,R_\mathrm{max}}$ (crosses). The relation can be well expressed by a linear function and implies that $M_\mathrm{thres}$ depends on two EoS parameters namely $M_\mathrm{max}$ and $R_\mathrm{max}$. The ratio $M_\mathrm{thres}/M_\mathrm{max}$ can be similarly described by a quantity $C^{*}_{1.6}=\frac{G\,M_\mathrm{max}}{c^2\,R_{1.6}}$, which does not have a direct physical meaning but which does involve the radius $R_{1.6}$ of a nonrotating NS with 1.6~$M_\odot$ (circles in Fig.~\ref{fig:kcmax}). As described in Sec.~\ref{dominant}, $R_{1.6}$ can be accurately measured for instance by postmerger GW emission from 1.35-1.35~$M_\odot$ binaries. If $R_{1.6}$ was measured sufficiently well, $k$ and thus $M_\mathrm{thres}$ depend on $M_\mathrm{max}$ only. A measurement of $M_\mathrm{thres}$ or a constraint on this quantity can then be employed to determine the unknown maximum mass $M_\mathrm{max}$ of nonrotating NSs.

The threshold binary mass $M_\mathrm{thres}$ can be measured or at least constrained by observing several NS mergers, whose total binary masses are inferred from the GW inspiral signal and whose merger outcome can be assessed through either the detection of postmerger GW emission or through the properties of the electromagnetic counterpart. Determining $M_\mathrm{max}$ is particularly interesting because this quantity depends on the very high-density regime which may not be directly accessible by other measurements or observations of NSs in the canonical mass range around 1.4~$M_\odot$ \cite{Oezel2016}. We stress that already a single detection of an event with clear indications of a prompt collapse could yield an upper limit on $M_\mathrm{max}$ even if $R_{1.6}$ is not yet well constrained \cite{Bauswein2013,Bauswein2016}. An observational upper limit on $M_\mathrm{max}$ may be difficult to obtain otherwise, as for instance any pulsar observation in double NS systems can only provide a lower bound on $M_\mathrm{max}$. In this context, we refer to some recent studies in connection with GW170817 or short gamma-ray burst observations, which consider the stability of a potential uniformly rotating late-time remnant to place constraints on $M_\mathrm{max}$ e.g.~\cite{Lasky2014,Lawrence2015,Fryer2015,Margalit2017,Shibata2017,Rezzolla2018,Ruiz2018}. We remark that these conclusions rely on certain interpretations of the electromagnetic emission and on further assumptions about the properties of the late-time remnant, which are hard to check within models.

In Sec.~\ref{dominant}. we already described that the dominant postmerger GW frequency $f_\mathrm{peak}^\mathrm{thres}$ of a system with a total binary mass $M_\mathrm{stab}$ (so slightly below $M_\mathrm{thres}$), scales tightly with the radius $R_\mathrm{max}$ of the maximum-mass configuration of nonrotating NSs. Similarly, $f_\mathrm{peak}^\mathrm{thres}$ correlates with the maximum density of nonrotating NSs, as was found in \cite{Bauswein2013} and \cite{Bauswein2014a}. This stresses the importance of measuring $f_\mathrm{peak}^\mathrm{thres}$ and $M_\mathrm{thres}$. We recall here the main idea of \cite{Bauswein2014a} namely that $M_\mathrm{stab}$ (or equivalently $M_\mathrm{thres}$\footnote{Note the slightly different nomenclature concerning $M_\mathrm{thres}$ and $f_\mathrm{thres}$ in \cite{Bauswein2013} and \cite{Bauswein2014a}.}) and $f_\mathrm{peak}^\mathrm{thres}$ can be estimated through an extrapolation procedure from at least two measurements of the dominant postmerger GW frequency of events with lower and distinct total binary masses, which are within the most likely range of $M_\mathrm{tot}$ (see Fig.  \ref{fig:rmboxes}). A determination or estimate of $M_\mathrm{thres}$ and $f_\mathrm{thres}$ will provide valuable insights into the properties of the EoS at very high densities.

\section{Semi-analytic model for the threshold binary mass}\label{sec:rns}

The ideas to constrain properties of high-density matter as laid out in this review rely on empirical relations between observables, e.g.\ GW frequencies, and characteristic quantities of the EoS, usually stellar properties of nonrotating NSs. Many of these empirical relations are intuitive and in this sense expected and understandable, e.g.\ the dependence of the dominant postmerger GW frequency on NS radii. Other relations may be less obvious and may be seen as a mere outcome of simulations. In principle, selection effects could be introduced by the methodology of the simulation tool or the choice of candidate EoSs. It is therefore important to corroborate the existence of the respective scalings. One of the important, but less intuitive, relations is the dependence of $k=M_\mathrm{thres}/M_\mathrm{max}$ on $C_\mathrm{max}$ as shown in the left panel of Fig.~\ref{fig:kcmax} (but see \cite{Bauswein2013} for a simplistic model assuming polytropic EoSs and Newtonian physics).

We therefore developed a semi-analytic model, which confirms the robustness of the relation shown in Fig.~\ref{fig:kcmax}. The study still adopts certain simplifications, but is largely independent from hydrodynamical simulations and considers a large class of EoSs. In \cite{Bauswein2017a} we construct relativistic stellar equilibrium configurations of differentially rotating NSs with a prescribed rotation law using the publicly available RNS code \cite{Stergioulas1995}. A merger remnant is modeled by an equilibrium configuration, with the corresponding mass and angular momentum. By analyzing sequences of models, one can determine the maximum mass for a given amount of angular momentum. This can be compared to the available angular momentum in a merger remnant of a given total binary mass, which is a very robust result from simulations. Specificallly, for each EoS we find a linear empirical relation of the form \cite{Bauswein2017a}
\begin{equation}
J _ { \rm { merger } } \simeq a M _ { \mathrm { tot } } - b
\end{equation}
for the angular momentum in the remnant $J _ { \rm { merger }}$ as a function of the total mass $M _ { \mathrm { tot }}$. For example, $a=4.041$ and $b=4.658$ for the DD2 EoS.

The comparison reveals whether a merger remnant of a given mass possessed sufficient angular momentum to be stable against a gravitational collapse. Within this simplified, but semi-analytic, model one can compute a theoretical threshold binary mass for a given EoS. In the right panel of Fig.~\ref{fig:kcmax} this estimated threshold mass (crosses) is compared to the threshold binary mass, which is determined from hydrodynamical simulations (solid line shows a least-squares fit to the results from simulations, i.e.\ the data in the left panel of Fig.~\ref{fig:kcmax}). The theoretical estimate of $M_\mathrm{thres}$ agrees to within 3-7 per cent with the true threshold mass. In Fig.~\ref{fig:kcmax} (right panel) one recognizes a slight underestimation, but the same qualitative behavior, which strongly supports the existence of a tight relation between $k=M_\mathrm{thres}/M_\mathrm{max}$ and $C_\mathrm{max}$. A perfect quantitative agreement may not be expected, because the semi-analytic model adopts zero-temperature EoSs and an ad-hoc choice of the rotation law, apart from other assumptions, for instance that the dynamical, early merger phase can be at all described by equilibrium models (see \cite{Bauswein2017a} for details).

\section{Neutron star radius constraints from GW170817}\label{sec:gw17}

The corroboration of the particular EoS dependence of $M_\mathrm{thres}$, as displayed in Fig.~\ref{fig:kcmax}, is finally important for a first direct application of these relations in connection with GW170817, in order to constrain NS radii \cite{Bauswein2017}. A least square fit to the relation between $k=M_\mathrm{thres}/M_\mathrm{max}$ and $C_\mathrm{max}=\frac{G\,M_\mathrm{max}}{c^2\,R_\mathrm{max}}$ (Fig.~\ref{fig:kcmax}) yields
\begin{equation}\label{eq:mthr}
M_\mathrm{thres}=\left( -3.38\frac{G\,M_\mathrm{max}}{c^2\,R_\mathrm{max}} + 2.43 \right)\,M_\mathrm{max}.
\end{equation}
Moreover, $M_\mathrm{max}$ and $R_\mathrm{max}$ are not completely uncorrelated. Causality requires $M_\mathrm{max}\leq\frac{1}{2.823}\frac{c^2\,R_\mathrm{max}}{G}$ because the speed of sound is limited by the speed of light, which implies that an EoS cannot become arbitrarily stiff (see \cite{Koranda1997,Lattimer2016} for more details). Inserting this requirement in Eq.~(\ref{eq:mthr}) results in the constraint
\begin{equation}\label{eq:radcon}
M_\mathrm{thres}\leq\left( -3.38\frac{1}{2.823} + 2.43 \right) \frac{1}{2.823}\frac{c^2\,R_\mathrm{max}}{G}=0.436\frac{c^2\,R_\mathrm{max}}{G}.
\end{equation}
Hence, a given measurement or estimate of $M_\mathrm{thres}$ sets a lower bound on $R_\mathrm{max}$. 

The total binary mass measured in GW170817 was originally reported as $2.74^{+0.04}_{-0.01}~M_\odot$ in \cite{Abbott2017}\footnote{The total binary mass of GW170817 was slightly revised to $2.73^{+0.04}_{-0.01}~M_\odot$ in \cite{Abbott2019}.}. The ejecta mass in this event has been estimated from the properties of the electromagnetic counterpart to be in the range of 0.03 to 0.05~$M_\odot$ \cite{Cowperthwaite2017,Kasen2017,Nicholl2017,Chornock2017,Drout2017,Smartt2017,Kasliwal2017,Kilpatrick2017,Perego2017,Tanvir2017,Tanaka2017}. Although these ejecta mass estimates involve some uncertainties, the amount of unbound matter in GW170817 is certainly at the high end of what is expected from numerical merger simulations for any EoS. Based on this observation, we argued in \cite{Bauswein2017} that the high ejecta mass in GW170817 strongly suggests that the merger did not result in a prompt collapse, because direct BH formation implies significantly reduced mass ejection (see e.g. Fig.~7 in \cite{Bauswein2013a}). If this hypothesis is correct, the measured total binary mass of GW170817 is smaller than the threshold binary mass $M_\mathrm{thres}$ for prompt BH formation and thus $M_\mathrm{thres}\geq 2.74^{+0.04}_{-0.01}~M_\odot$. Using this condition in Eq.~(\ref{eq:radcon}) results in a lower limit on $R_\mathrm{max}$.

     \begin{figure*}
\begin{center}
\resizebox{0.9\textwidth}{!}{
\includegraphics{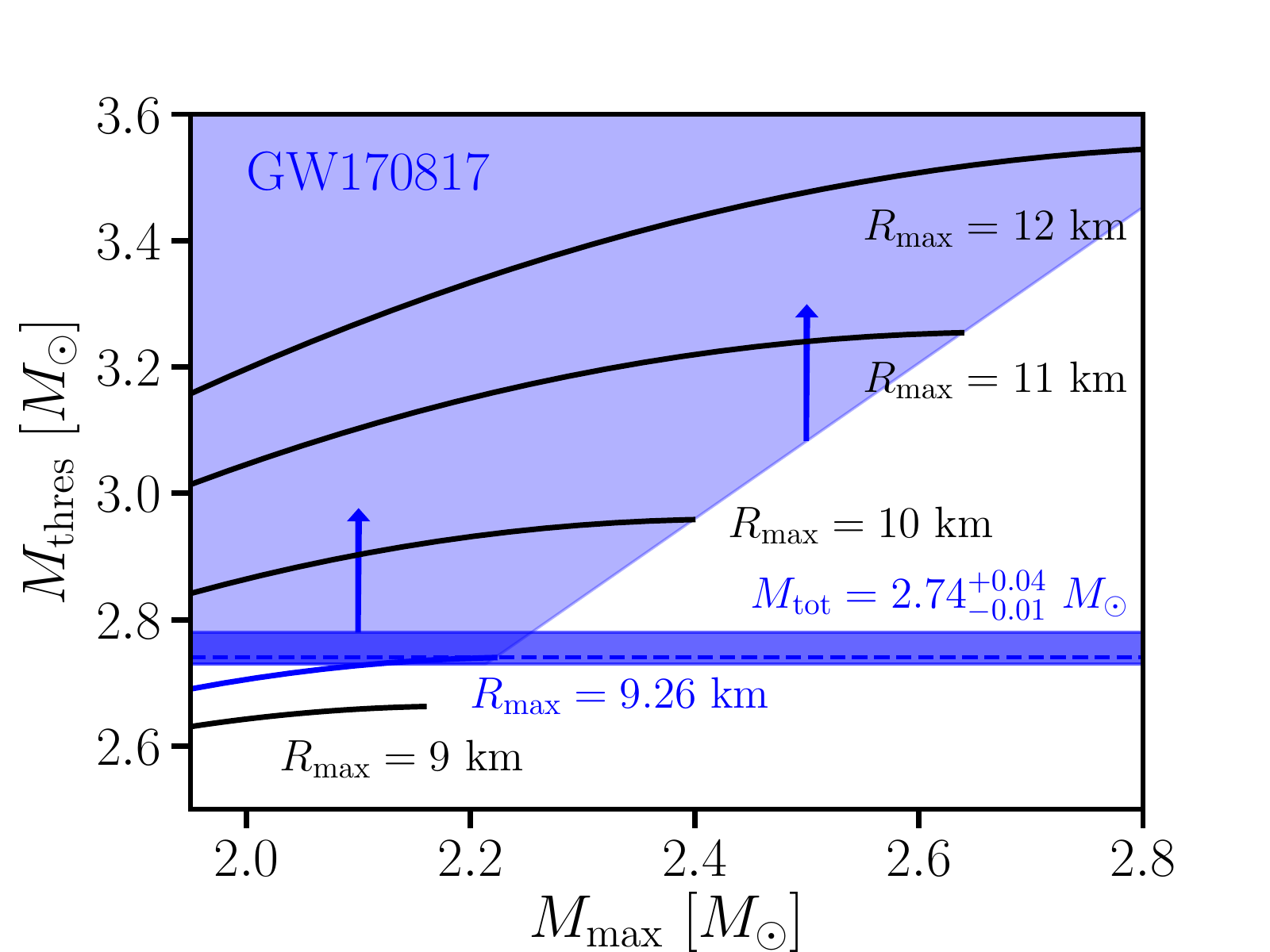}
}
\end{center}
  \caption{Threshold binary mass $M_\mathrm{thres}$ for prompt collapse as a function of $M_\mathrm{max}$ for different $R_\mathrm{max}$ (solid lines). The dark blue band shows the total binary mass of GW170817, providing a lower limit on $M_\mathrm{thres}$. The true $M_\mathrm{thres}$ must lie within the light blue areas if GW170817 resulted in a delayed/no collapse. This rules out NSs with $R_\mathrm{max}\leq9.26^{+0.17}_{-0.03}$~km. Figure from \cite{Bauswein2017}.}
  \label{fig:mthmmax}
\end{figure*}

These arguments are visualized in Fig.~\ref{fig:mthmmax}, which shows $M_\mathrm{thres}$ (solid lines) as function of $M_\mathrm{max}$ for different fixed radii $R_\mathrm{max}$. The dark blue band indicates the measured total binary mass of GW170817. The true $M_\mathrm{thres}$ must lie above the lower edge of the dark blue band if $M_\mathrm{thres}\geq 2.74^{+0.04}_{-0.01}~M_\odot$. Moreover, causality limits the allowed values of $M_\mathrm{thres}$ and $M_\mathrm{max}$ to the upper left corner of this figure. Small radii are incompatible with $M_\mathrm{thres}\geq 2.74^{+0.04}_{-0.01}~M_\odot$, because they do not yield sufficiently high threshold masses.

\begin{figure*}
\begin{center}
\resizebox{1\textwidth}{!}{
\includegraphics{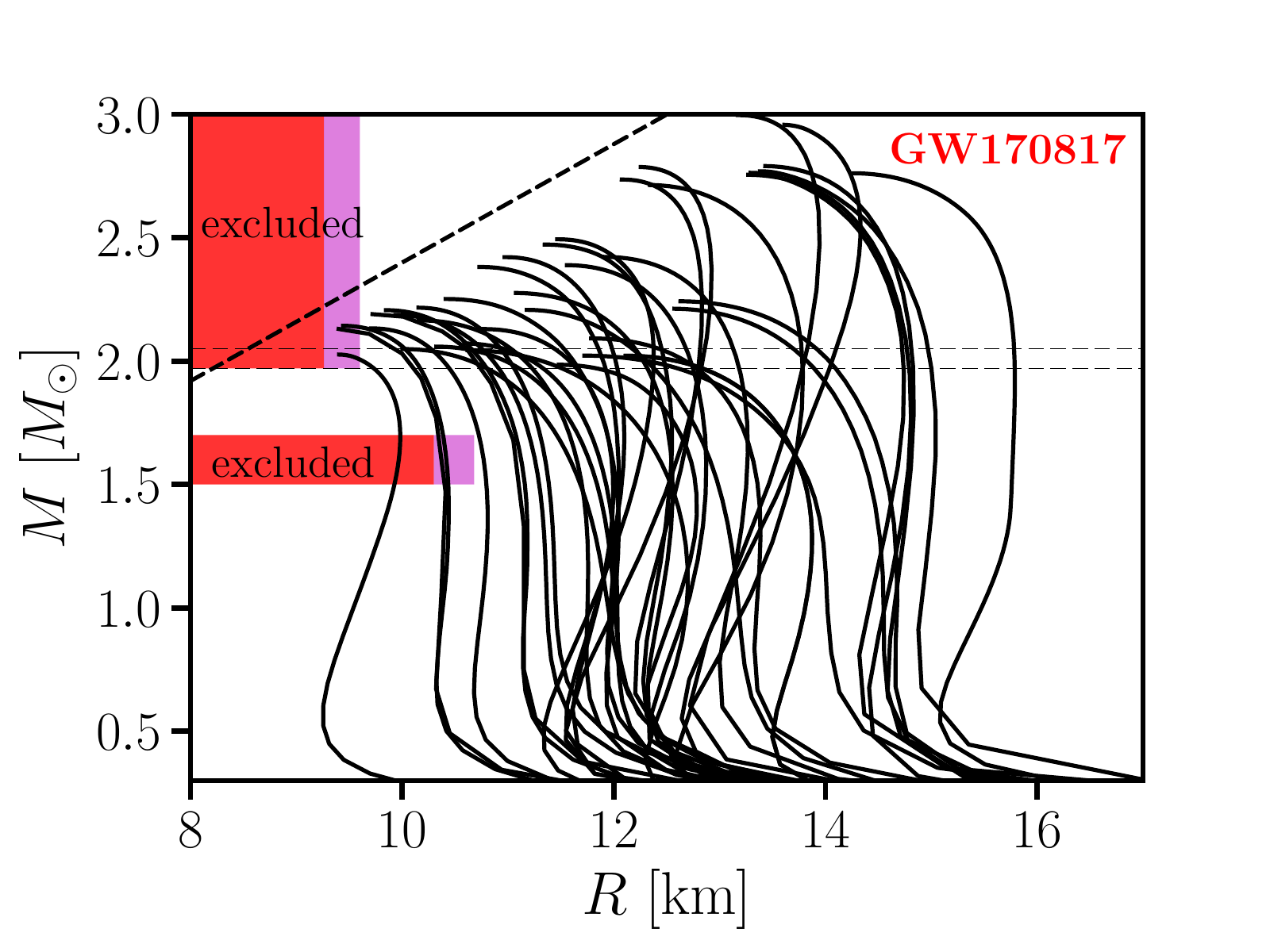}
\includegraphics{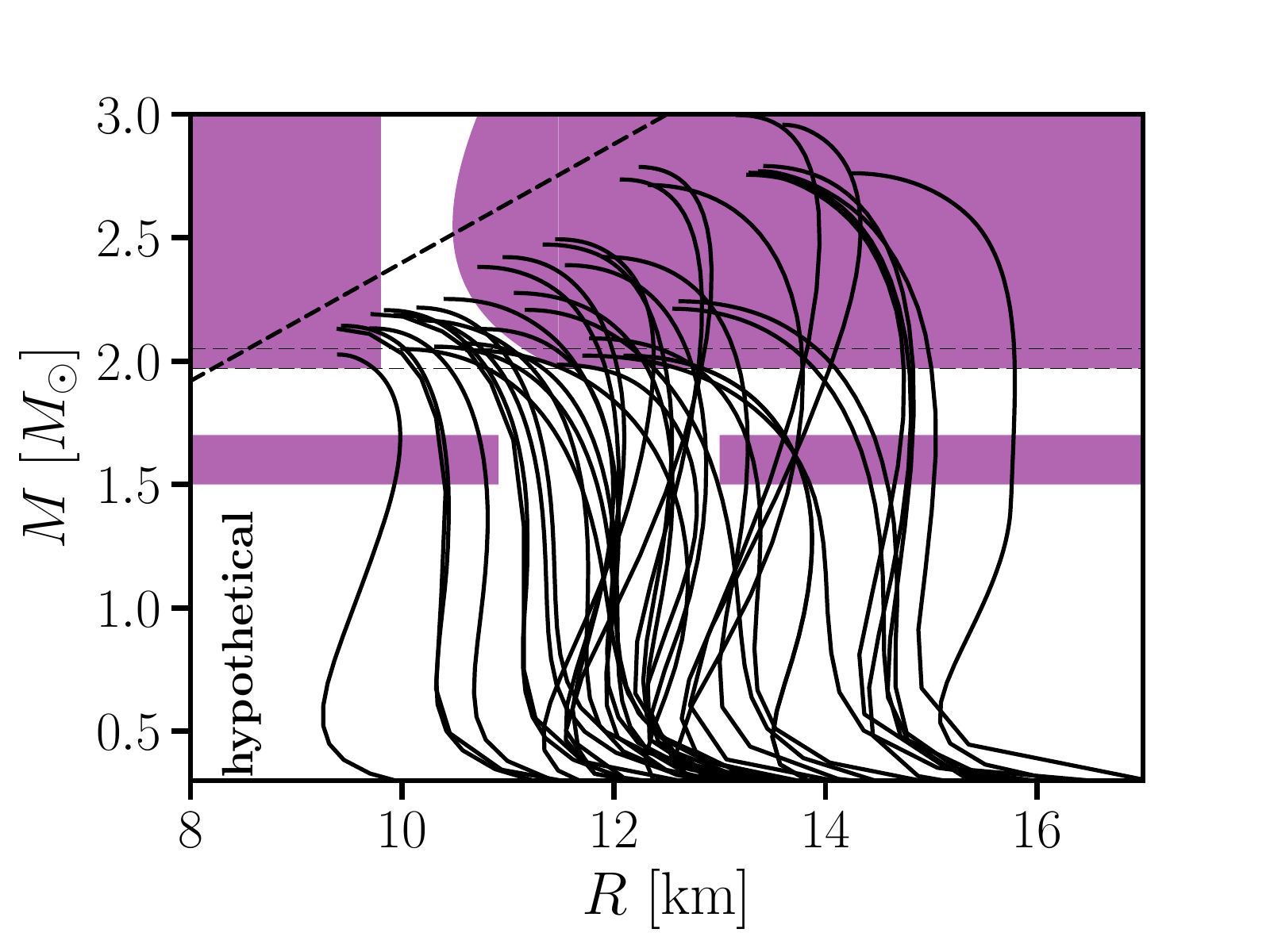}
}
\end{center}
  \caption{Left panel: Mass-radius relations of different EoSs with very conservative (red area) and ``realistic'' (cyan area) constraints derived from the measured total binary mass of GW170817 under the assumption of no prompt BH formation of the merger remnant (see \cite{Bauswein2017} for details). Horizontal lines display the limit by \cite{Antoniadis2013}. The thick dashed line shows the causality limit. Right panel: Mass-radius relations of different EoSs with hypothetical exclusion regions (purple areas) from a delayed-collapse event with $M_\mathrm{tot}=2.9~M_\odot$ and a prompt-collapse event with $M_\mathrm{tot}=3.1~M_\odot$ employing arguments based on the collapse behavior. Figures from \cite{Bauswein2017}.}
  \label{fig:radcon}
\end{figure*}

Refining the argumentation the detailed calculation and error analysis in \cite{Bauswein2017} yields $R_\mathrm{max}\geq9.60^{+0.14}_{-0.03}$~km. Following the same line of arguments, but using the relation $k=k(M_\mathrm{max}/R_{1.6})$ (circles in Fig.~\ref{fig:kcmax}), which equivalently describes $M_\mathrm{thres}$ as function of $M_\mathrm{max}$ and $R_{1.6}$, provides a lower limit on the radius $R_{1.6}$ of a nonrotating NS with 1.6~$M_\odot$. The radius $R_{1.6}$ has to be larger than $10.68^{+0.15}_{-0.04}$~km. These NS radius constraints are displayed in the left panel of Fig.~\ref{fig:radcon} on top of a set of mass-radius relations of EoSs which are available in the literature (see \cite{Bauswein2012a,Bauswein2013a} for an overview of these EoS models). Our method, which is based on a set of minimal assumption, places an absolute lower limit to NS radii that clearly rules out very soft nuclear matter (we note that all these results are derived in the framework of general relativity and different lower limits would apply in alternative theories of gravity). Ref. \cite{Radice2018} follows a very similar idea to find a lower limit on the tidal deformability of NSs. We also refer to the analysis in \cite{Abbott2017,De2018,PhysRevLett.121.161101}, which excludes very stiff EoSs through an upper limit on the tidal deformability, which can be converted to an upper limit of NS radii of about 14~km, e.g.~\cite{Fattoyev2018,Raithel2018}. A more detailed analysis in \cite{PhysRevLett.121.161101} arrives at the range $10.5 {\rm km} \leq R _ { 1.4 } \leq 13.3{\rm km}$ for the radius of $1.4M\odot$ NSs at a 90\% credible level. For additional estimates of NS radii based on the observation of GW170817, see Table II in \cite{2018arXiv181110929M} and references therein.

\begin{figure*}
\begin{center}
\resizebox{0.9\textwidth}{!}{
\includegraphics{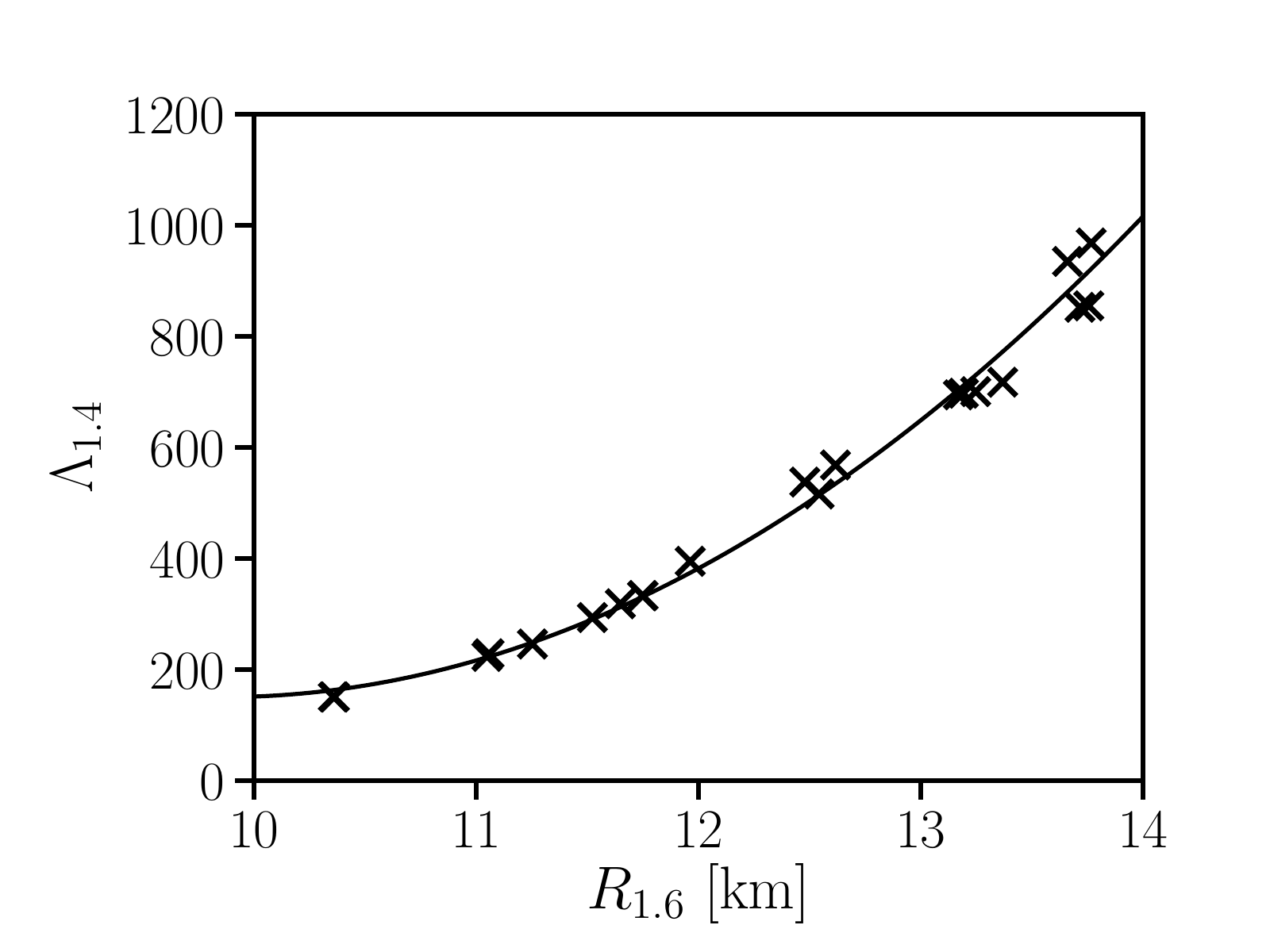}
}
\end{center}
  \caption{Tidal deformability of a 1.4~$M_\odot$ NS as function of radius of a 1.6~$M_\odot$ NS for different EoSs. Solid line shows a quadratic least-square fit.}
  \label{fig:lamrad}
\end{figure*}

Considering the tight scaling of NS radii with the tidal deformability it is straightforward to convert our radius constraints to limits on the tidal deformability. A limit of $R_{1.6}>10.7$~km corresponds to a lower bound on the tidal deformability of a 1.4~$M_\odot$ NS of about $\Lambda_{1.4}>200$.  This can be seen in Fig.~\ref{fig:lamrad} showing the tidal deformability of a 1.4~$M_\odot$ NS as function of $R_{1.6}$ for many different EoS models. The empirical relation allows a approximate conversion of both quantities.

Throughout the derivation we made conservative assumptions and considered the corresponding error bars, which is why our radius constraint and the equivalent limit on the tidal deformability are less strong but more robust compared to similar constraints in the literature~\cite{Radice2018,Coughlin2018,Most2018,Radice2018a,Coughlin2018a}, some of which rely on analogous arguments. We emphasize that the lower limit on the radius and the tidal deformability cannot be significantly higher than our bound, since otherwise conflicts with existing literature data arise. There are for instance simulations for EoSs with only somewhat larger radii and tidal deformability that do not result in a prompt collapse and would thus presumably lead to a relatively bright electromagnetic counterpart compatible with GW170817. 

We remind that it is precisely the threshold between dim and bright electromagnetic counterparts that our derivation aims to determine in a conservative way. Hence, the resulting limits cannot be much different if arguments about the brightness of the electromagnetic transient are employed.

Although these current radius constraints are not yet very restrictive, we highlight two aspects of our new method. First, our method is very robust and conservative. Apart from the hypothesis of no direct collapse in GW170817, all errors can be quantified and we make conservative assumptions throughout the analysis. This robustness is a particular advantage of our constraints. The only required result from simulations is the particular EoS dependence of $M_\mathrm{thres}$. This theoretical input can be tested by more elaborated simulations in the future, but it seems unlikely that for instance a detailed incorporation of neutrinos or magnetic fields can have a significant impact on the threshold binary mass. For instance, these effects have only a weak impact on the dominant postmerger GW emission (e.g.~\cite{Sekiguchi2011a,Kiuchi2014,Palenzuela2015,Foucart2016,Endrizzi2016,Kawamura2016}), which is why one may expect also a subdominant influence on other bulk features of NS mergers, such as the threshold mass. 
As described above, the qualitative behavior of the EoS dependence of $M_\mathrm{thres}$ has been confirmed by a robust semi-analytic model. Moreover, the assumption of no prompt collapse can be verified by more refined merger and emission models in the future. Also, as more events are observed in the future our understanding of kilonovae will grow and the interpretation of the emission properties and the underlying merger dynamics will become more obvious. Second, it is expected that more events similar to GW170817, i.e. events which allow a (tentative) distinction between prompt and no direct collapse of the remnant in combination with a precise measurement of the total binary mass, will be observed in the near future. We then expect to obtain more stringent constraints on $M_\mathrm{thres}$ in combination with a more robust classification of the collapse behavior through the electromagnetic counterpart. In particular, it is conceivable that a prompt-collapse event will set an upper bound on $M_\mathrm{thres}$. This will potentially limit the allowed stellar parameters to a very small parameter range (see e.g.\ Figs.~3 to~5 in \cite{Bauswein2017} for hypothetical future events). An upper limit on the threshold mass constrains the maximum mass and radii of nonrotating NSs from above. The resulting constraints from hypothetical future detections are shown in Fig.~\ref{fig:radcon} (right panel).

We stress that apart from its robustness, our method for EoS constraints has the advantage that it does not require GW detections with high SNRs. In contrast, direct measurements of EoS effects during the late inspiral phase or in the postmerger stage both rely on a detection of a very strong signal. It is very likely that EoS constraints through the collapse behavior will soon provide more detailed insights into the properties of high-density matter.

The work described in this review highlights several aspects of NS mergers, which link them to fundamental questions of physics. This includes the formation of r-process elements and properties of high-density matter. The era of multi-messenger observations of NS mergers has just begun with the first detection of GWs from a NS coalescence and accompanying electromagnetic emission. More of such observations can be expected in the near future. The ideas outlined above have emphasized the scientific value of such measurements, the interpretation of which, however, critically relies on simulations of the merger process. 

Improvements of the simulation tool will lead to more reliable predictions of the anticipated GW signals concerning, for instance, the damping of postmerger waveforms. This is important to advance current GW data analysis methods for the measurement of postmerger GW features as discussed in Sec.~\ref{dominant} and ~Sec. \ref{secondary}. Our work focuses on the three most prominent peaks in the postmerger GW spectrum. It will be useful to also comprehend the physical mechanisms behind some other secondary features of the postmerger phase. Understanding their dependencies on the EoS may reveal further details of the properties of high-density matter. Such an analysis may be complemented by theoretical models of the postmerger remnant, which are based on perturbative calculations that are successfully applied to oscillations of isolated NSs.

\section{Summary}\label{sec:sum}
In this work, we summarized several aspects which are related to the postmerger phase of NS mergers. We focus on the GW emission after merging and on the immediate outcome of the collision, i.e. the distinction between the formation of a BH or of a NS remnant. The main objective of these considerations is setting constraints on the incompletely known EoS of high-density matter, where we focus on methods complementary to the extraction of finite-size effects during the inspiral phase before merging.

The most remarkable feature of the postmerger GW emission is a pronounced peak in the spectrum between roughly 2~kHz and 4~kHz. This peak robustly occurs in all models which do not promptly form a BH. The peak reflects the dominant oscillation mode of the remnant which is excited by the merging process. The peak depends sensitively on the total binary mass and the stiffness of the adopted EoS. The mass ratio has a weaker impact on the position of the peak.

For fixed individual binary masses the peak frequency scales tightly with radii of nonrotating NSs with a fixed fiducial mass. In turn, this offers the possibility to infer NS radii from a measurement of the dominant postmerger GW frequency. First studies with simulated injections have revealed that such observations can possibly be achieved for events at a few ten Mpc with current detectors at design sensitivity or with projected upgrades to the available instruments. For these detections a precise measurement of the binary mass ratio is less relevant, since the frequency vs. radius relation exists also for systems with constant chirp mass, which can be extracted from observations with high precision. Ongoing research should further explore GW data analysis strategies and refine existing hydrodynamical simulations of the merger process since the precision of the extracted EoS/radius information relies on empirical relations, which are determined through numerical models.

Remarkably, the dominant postmerger GW frequency may also indicate the occurrence of deconfined quark matter in the merger remnant through a strong first order phase transition. A significant increase of the peak frequency relative to the tidal deformability, which is measured from the inspiral phase, provides an unambiguous signature of a strong phase transition. This nicely illustrates the complementary information which can be inferred from the postmerger and the inspiral phases. The reason is that the two phases probe different density regimes of the NS EoS, where the dynamics of the postmerger stage are sensitive to the very high density regime, since densities increase after merging.

Apart from the main peak, the postmerger GW spectrum contains additional subdominant peaks, which encode information on the structure and dynamics of the remnant. Observationally, only those with frequencies below the main peak are relevant, because the sensitivity of ground-based GW detectors degrades at higher frequencies. Not all mechanisms and oscillation modes generating secondary features are identified yet. But, there is good evidence that the two most pronounced secondary peaks are produced by a non-linear coupling of the quasi-radial mode with the dominant oscillation and by the formation of tidal bulges at the remnant surface during merging, respectively. It is important to note that, depending on the total binary mass and the EoS, these features appear with different strength or can even be absent. 

This leads to three different morphologies of the GW spectrum, depending on whether one or the other or both secondary features are present with an appreciable amplitude. The three different types of the spectral classification scheme are also reflected in different dynamical behaviors during merging and the early remnant evolution. This is an important, although not surprising, connection between the observable GW signal and the underlying dynamics of the merger process. Detecting secondary features can support the inference of EoS information from postmerger GW spectra. Future work should clarify the nature and origin of other subdominant features in the GW sepctrum. Also, the impact of the binary mass ratio on secondary features is not yet fully explored.

These discussions highlight the potential of future GW detections with increased sensitivity in the kHz range. The detection of postmerger GW emission represents a highly rewarding target to understand the properties of high-density matter (independent of and complementary to finite-size effects during the inspiral) and the dynamics of the early postmerger remnant evolution, which can be linked to the electromagnetic emission within a multi-messenger picture of NS mergers. The scenarios discussed above underline the outstanding importance to install upgrades to the current generation of ground-based GW detectors with a better sensitivity at high frequencies and to develop the next generation of instruments.

This article also summarizes the dependencies of the gravitational collapse during merging. Central is the threshold total binary mass for direct BH formation. Systems with total masses above the threshold mass lead to a prompt collapse, whereas less massive binaries result in the formation of a massive, rotating NS remnant. Importantly, the different outcomes can be observationally distinguished, e.g. by the presence of postmerger GW emission at a few kHz or the detailed properties of the electromagnetic transient produced by radioactive decays in the ejecta.

It is found empirically that the threshold mass is well described by the maximum mass $M_\mathrm{max}$ of nonrotating NSs and the radius of nonrotating NSs. This dependence is corroborated by a semi-analytic model employing a stellar equilibrium code. In future, the particular dependence of the threshold mass on NS properties offers the possibility to determine $M_\mathrm{max}$ if NS radii are known with sufficient precision. The threshold mass can be determined by a number of GW detections that reveal the total binary masses and that allow an observational distinction between prompt and no direct collapse, e.g. through postmerger GW emission or through the properties of the electromanetic counterpart. More detailed studies are required to interpret the quasi-thermal emission of radioactively powered ejecta and to robustly infer underlying ejecta properties and the merger outcome. This includes highly resolved hydrodynamical models and radiative transfer calculations.

For GW170817, the relatively high inferred ejecta mass may provide tentative evidence that the merger remnant did not directly collapse into a BH, because this would likely result in a dimmer emission in the infrared and optical. This interpretation sets a lower bound on NS radii of typical mass of about 10.7~km, or alternatively a bound on the tidal deformability. The reason is that EoSs resulting in smaller radii would inevitable lead to a prompt collapse independent of the maximum mass of nonrotating NSs. Thus, the multi-messenger interpretation of GW170817 rules out very soft nuclear matter based on a minimum number of assumptions. This lower bound on NS radii complements the upper limit derived from constraints of the tidal deformability resulting from the GW inspiral phase.

This type of considerations bear a lot of potential for the near future as more GW events with accompanying electromagnetic emission are observed. Future NS merger observations will likely reveal different total binary masses and will elucidate the possible variation in the electromagnetic radiation. A sample of events will lead to a better understanding of the electromagnetic emission processes and the underlying merger dynamics. Hence, a more robust interpretation and inference of the merger outcome will be possible. This will result in stronger lower bounds on NS radii. If good evidence for a prompt-collapse event is found, upper limits on NS radii and the maximum mass of nonrotating NSs can be established by the same method.

\section*{Acknowledgments}

AB acknowledges support by the European Research Council (ERC) under the 
European Union's Horizon 2020 research and innovation programme under 
grant agreement No. 759253, the Klaus-Tschira Foundation and the German Research Foundation (DFG) via the Collaborative Research Center SFB 881 ‘The Milky Way System’ and SFB 1245 ‘Nuclei: From Fundamental Interactions to Structure and Stars’. NS acknowledges support by the COST Actions GWVerse (CA16104), G2Net (CA17137) and PHAROS (CA16214) and by the ARIS facility of GRNET in Athens (GWAVES and GRAVASYM allocations).

\bibliographystyle{iopart-num}
\bibliography{thesis}

\end{document}